\documentclass[lettersize,journal]{IEEEtran}

\usepackage{amsmath,amsfonts}
\usepackage{algorithmic}
\usepackage{algorithm}
\usepackage{array}
\usepackage[caption=false,font=normalsize,labelfont=sf,textfont=sf]{subfig}
\usepackage{textcomp}
\usepackage{stfloats}
\usepackage{url}
\usepackage{verbatim}
\usepackage{graphicx}
\usepackage{cite}
\usepackage{ulem}
\usepackage{multirow}  
\usepackage{rotating}
\usepackage{pifont} 

\def\BibTeX{{\rm B\kern-.05em{\sc i\kern-.025em b}\kern-.08em
    T\kern-.1667em\lower.7ex\hbox{E}\kern-.125emX}}

\begin{document}
\title{Reinforcement Learning-Empowered Mobile Edge Computing for 6G Edge Intelligence}
\author{Peng~Wei,
Kun~Guo, \IEEEmembership{Member, IEEE},
Ye~Li, \IEEEmembership{Member, IEEE},
Jue~Wang, \IEEEmembership{Member, IEEE}, \\
Wei~Feng, \IEEEmembership{Senior Member, IEEE},
Shi~Jin, \IEEEmembership{Senior Member, IEEE},
Ning~Ge, \IEEEmembership{Member, IEEE}, \\
and Ying-Chang~Liang, \IEEEmembership{Fellow, IEEE}%
\thanks{

Peng Wei, Wei Feng, and Ning Ge are with the Beijing National Research Center for Information Science and Technology, Department of Electronic Engineering, Tsinghua University, Beijing 100084, China (e-mail: wpwwwhttp@163.com; fengwei@tsinghua.edu.cn; gening@tsinghua.edu.cn).


Kun Guo is with the School of Communications and Electronics Engineering, East China Normal University, Shanghai 200241, China (e-mail: kguo@cee.ecnu.edu.cn).

Ye Li and Jue Wang are with the School of Information Science and Technology, Nantong University, Nantong 226019, China (e-mail: yeli@ntu.edu.cn; wangjue@ntu.edu.cn).

Shi Jin is with the National Mobile Communications Research Laboratory, Southeast University, Nanjing 210096, China (e-mail: jinshi@seu.edu.cn).



Ying-Chang Liang is with the National Key Laboratory on Communications, Center for Intelligent Networking and Communications, University of Electronic Science and Technology of China, Chengdu 611731, China (e-mail: liangyc@ieee.org).
}
}


{}
\maketitle

\begin{abstract}
Mobile edge computing (MEC) is considered a novel paradigm for computation-intensive and delay-sensitive tasks in fifth generation (5G) networks and beyond.
However, its uncertainty, referred to as dynamic and randomness, from the mobile device, wireless channel, and edge network sides, results in high-dimensional, nonconvex, nonlinear, and NP-hard optimization problems.
Thanks to the evolved reinforcement learning (RL), upon iteratively interacting with the dynamic and random environment, its trained agent can intelligently obtain the optimal policy in MEC.
Furthermore, its evolved versions, such as deep RL (DRL), can achieve higher convergence speed efficiency and learning accuracy based on the parametric approximation for the large-scale state-action space.
This paper provides a comprehensive research review on RL-enabled MEC and offers insight for development in this area.
More importantly, associated with free mobility, dynamic channels, and distributed services, the MEC challenges that can be solved by different kinds of RL algorithms are identified, followed by how they can be solved by RL solutions in diverse mobile applications.
Finally, the open challenges are discussed to provide helpful guidance for future research in RL training and learning MEC.
\end{abstract}


\section{Introduction}
\begin{figure*}[!t]
\centering
\includegraphics[width=6in]{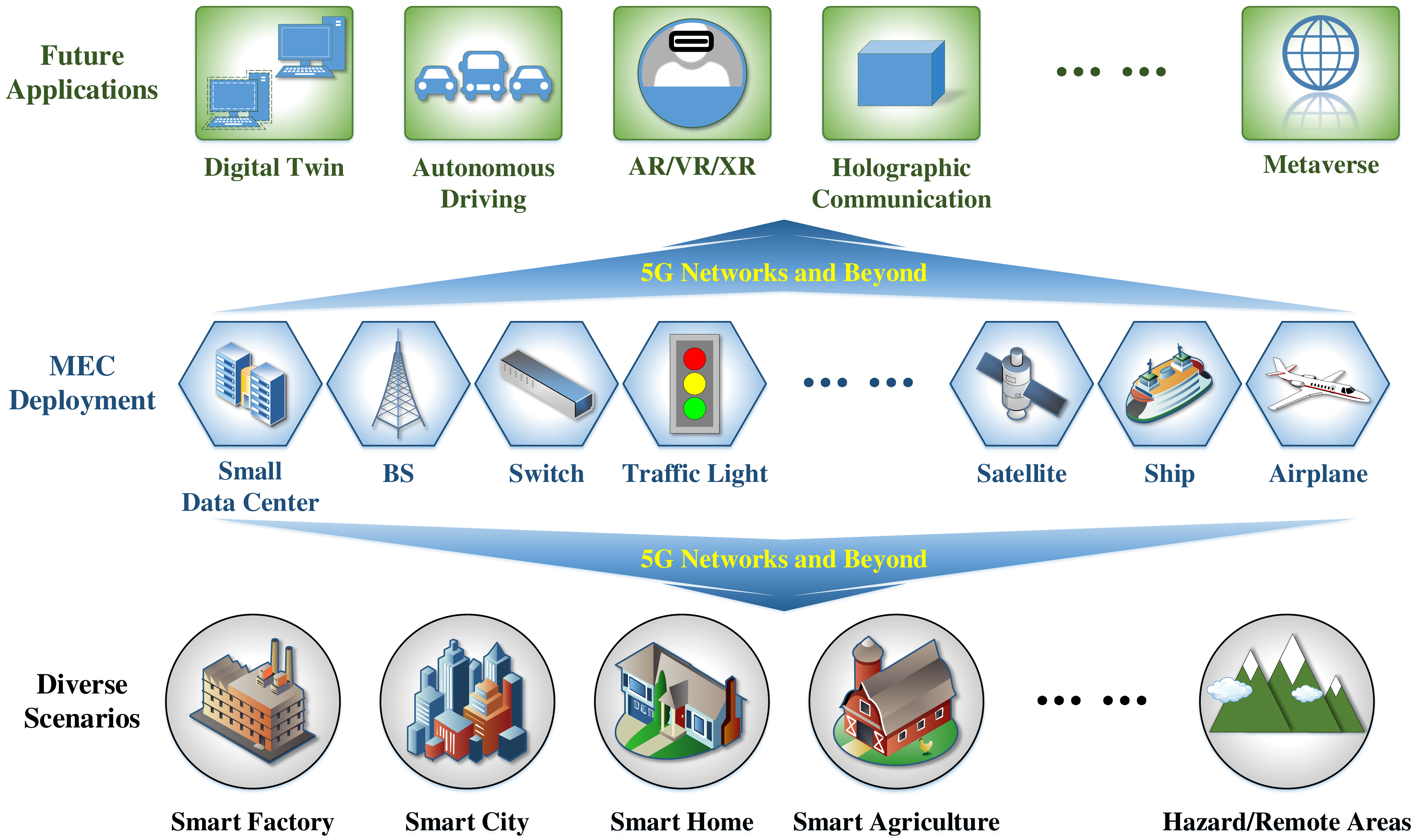}
\caption{Overview of MEC-empowered future networks.}
\label{newfig:fig1}
\end{figure*}

Witnessing the massive deployments of the service-oriented fifth generation (5G) networks, intelligence-oriented sixth generation (6G) is increasingly researched to facilitate the fully connected world \cite{9453860,9385374}, especially for edge intelligence \cite{9311932,9193946,9382020}.
It provides an emerging integrated communication network, artificial intelligence, and mobile edge computing (MEC) framework.
In recent years, developed based on previous edge computing technologies \cite{MicroDatacenterRef1,Ref7,FogRef1,Ref3}, MEC has become a new paradigm by shifting some centralized mobile cloud computing (MCC) functions in close proximity to edge devices within the radio access network (RAN) \cite{Ref8,Ref9,Ref47,Ref10}.
It provides highly efficient services for computation-intensive or delay-sensitive tasks by flexibly handling the compromise between uploading data to a central cloud server and processing data in capability-constrained mobile devices.
Compared to MCC \cite{Ref4,Ref5}, MEC provides a promising solution for supporting capability-constrained mobile devices accessing the network through resource-limited wireless channels \cite{Ref6,Ref8,Ref9}, while practical issues such as long propagation delay, network congestion, infeasible connectivity, and privacy leakage can be well addressed by properly designing MEC schemes.
As the number of smart devices is dramatically increasing in diverse 5G and beyond network scenarios (e.g., smart factory, smart home, and smart city) shown in Fig. \ref{newfig:fig1}, with various applications such as virtual/augmented/extended reality (VR/AR/XR), digital twin, and autonomous driving \cite{Ref9,Ref10,Ref200,8869705}, MEC is envisioned to play an important role in satisfying the high-quality requirements of computation-intensive businesses \cite{Ref9,Ref10} and low-latency applications \cite{9197676,liu2021mecempowered}.

However, implementing MEC still has many challenges.
In a general cloud-edge-terminal architecture, integrating different edge devices (e.g., mobile phones, tablets, and ground vehicles), various resources (e.g., bandwidth, storage, and computation), and diverse terminal requests (e.g., copies, deletes, and links) increases the heterogeneity.
In addition, the dynamic changes in these edge devices, resources, and requests result in instability in MEC systems.
First, mobile devices may further increase the network dynamics, e.g., their free mobility will incur random task arrival, fast fading and unreliable wireless channels, and frequent handover among many service access points (APs).
Second, opportunistic wireless access may also result in inadequate bandwidth and severe interference issues.
When MEC servers are equipped in mobile devices, such as ground vehicles and unmanned aerial vehicles (UAVs), the network uncertainty is further increased.
Last, as opposed to centralized cloud servers, distributed MEC servers usually have much limited capabilities including computational resources, storage, and service coverage.
These factors make efficient MEC design in such a complex network generally a nontrivial problem.

To evaluate the MEC system performance, integrated optimization models of communication, computing, caching, and control have been formulated \cite{Ref10,Ref13}, where complex network optimization problems appear to be attributed to the following critical factors.
\begin{enumerate}
  \item {\em Large-dimensional and multiobjective problems.} In the formulated optimization problems, multiple objectives with many different constraints are always required, owing to the coupling of communication, computation, and caching resources.
      With the exponential growth of network devices, the optimization problem dimension and its solution further increase.
      As a result, the joint optimization problem is usually NP-hard.
  \item {\em Mixed-integer programming.} Mixed-integer programming usually appears in the MEC optimization model.
      For instance, for the typical issue of offloading computational tasks to MEC servers for rapid execution, i.e., task offloading, discrete offloading decision-making variables and continuous resource allocation variables should be jointly considered.
  \item {\em Nonlinearity.} Nonlinearity may lead to complex optimization problems, e.g., nonconvex problems.
      For example, when energy consumption is concerned, the computational resource allocation problem is always nonlinear since the computational energy is usually described by the third power of the central processing unit (CPU) frequency \cite{Ref10}.
  \item {\em Network dynamics and randomness.} More importantly, owing to the aforementioned network dynamics and randomness, it might be difficult to establish valid mathematical models in the problem formulation.
      For example, the consumption rates of time and energy can be difficult to describe \cite{Ref17}.
\end{enumerate}
The aforementioned things make the integrated optimization problems in MEC unable to be efficiently solved by conventional network optimization methods such as derivation-based, heuristic, and game-theoretical algorithms \cite{Ref17}.
Moreover, most of the conventional approaches only focus on one-shot optimization; therefore, it is difficult to adapt to dynamic environments to obtain satisfactory long-term performance \cite{Ref13,Ref15}.
Briefly, these problems that exist in conventional optimization of the MEC system are ``low-level intelligence".

To bring more ``high-level intelligence" into MEC systems to address these problems, machine learning (ML) frameworks, which can effectively exploit critical information in an unpredictable environment, have attracted significant interest in the design of wireless networks \cite{Ref11,Ref12,Ref13,Ref15,Ref16,Ref17,Ref18}.
Explicitly, a machine executes a given task by exploiting its experience to reliably enhance the specific performance, and conversely, it learns the execution with the exhibited performance to accumulate experiences.
Based on the learning pattern, there are three popular machine learning techniques: supervised learning, unsupervised learning, and reinforcement learning (RL) \cite{Ref18}.
\begin{itemize}
  \item In supervised learning \cite{Ref18,Ref20}, such as regression models, $K$-nearest neighbors, support vector machines, and Bayesian learning, with the aid of a knowledgeable external supervisor, it infers and models the distribution of the labeled training data to predict the output of each given input in terms of classification.
      It can be applied to channel estimation/detection, user location/behavior learning/classification, spectrum sensing/detection, etc.
      However, due to its high dependency on the labeled samples in the database, the MEC system performance of supervised learning may be significantly degraded in unpredictable environments.
  \item In contrast, without the supervisor, unsupervised learning first classifies the unlabeled data based on specific metrics, such as the Euclidean distance-based ``similarity metric".
      Then, it learns the system model by extracting the features of the output and the classified input and uncovering the hidden structure to improve a specific performance.
      Its typical algorithms are $K$-means clustering, principal component analysis, and independent component analysis, which can be used in small cell clustering, heterogeneous network clustering, smart grid user classification, etc.
      However, in unreliable radio network environments, the classification accuracy decreases, which readily causes slow and inaccurate actions in MEC systems.
  \item Different from the static solutions provided by supervised learning and unsupervised learning, RL gives a constantly evolving intelligent framework \cite{Ref11,Ref12,Ref13,Ref15,Ref16,Ref17,Ref18}.
      In RL \cite{Ref21}, an agent is enabled to make proper decisions based on frequent interactions with the stochastic and dynamic environment.
      Based on the Markov models, the RL operates in a feedback mechanism (a closed loop) without the knowledge of input data, where based on the previous and current states, the agent can execute its actions to maximize the reward function.
      Additionally, the above behaviors, including the states, actions, and rewards, accumulate to generate experiences.
      The classic RL algorithm is {\it Q}-learning, which suffers from the curse of dimensionality caused by the large dimensions of the state-action spaces.
      Accordingly, upon leveraging the low-dimensional representation for the high-dimensional state-action spaces in a deep neural network (DNN), deep reinforcement learning (DRL) is invented.
      Furthermore, to solve specific problems, such as large overestimations of action values and multimodal distribution of future rewards \cite{Ref16}, many advanced variants of DRL have been proposed, such as double DRL, distributed DRL, and rainbow DRL \cite{Ref16,Ref21}.
\end{itemize}

There are some surveys associated with RL in communications and networking, especially with DRL.
Specifically, Refs. \cite{Ref13} and \cite{Ref15} provide a comprehensive review of DRL in Internet of Things (IoT) networks.
Multiaccess edge computing is considered in the former, and autonomous IoT edge/fog/cloud computing systems are considered in the latter.
In \cite{Ref16}, DRL applications in communications and networking were summarized, where MEC-enabled caching and offloading were introduced.
Ref. \cite{Ref22} provides a selective overview of DRL-based MEC, where multiagent DRL-based offloading and caching were highlighted.
Surveys on DRL for mobile edge caching and autonomous driving networks were reviewed in \cite{Ref23} and \cite{Ref24}, respectively.
Focusing on the high dimensionality in MEC \cite{8847416}, a thorough survey of works using machine learning is given in terms of offloading decisions, resource allocation, server deployment, and overhead management.
In \cite{9507541}, DRL-based resource allocation in MEC was also investigated.
Furthermore, although the applications of conventional RL and DRL in MEC were surveyed in terms of task offloading and network caching \cite{Ref25}, several MEC system issues were not included, such as task scheduling and resource allocation.
Most of these existing works show how RL has been effectively applied to MEC-based networks due to network dynamics.
However, network uncertainty, including dynamics and randomness, is not comprehensively investigated in MEC.
For a specific uncertainty-induced MEC problem including several subproblems, the existing works do not clearly state how to introduce RL to solve the overall problem or its subproblems.
Additionally, the preferred RL strategies for solving the specific problem are not sufficiently introduced.
Thus, this paper provides a summary of the network uncertainty in the MEC in terms of free mobility, time-varying wireless channels, and distributed services.

To this end, our contributions of this paper are listed as follows:
\begin{itemize}
  \item The main issues of network uncertainty in MEC, i.e., the challenges that should be addressed and solved to guarantee high quality of service (QoS) and quality of experience (QoE), are categorized and explained.
      This is helpful not only for researchers interested in RL solutions but also for those working with MEC in 5G networks and beyond.
  \item Next, a general RL algorithm framework involving two uncertainty-focused strategies is provided to promote RL-enabled MEC as an important paradigm for future network orchestration and control.
  \item Corresponding to challenges and solutions, the recent and significant literature in academia and industry that utilizes RL with MEC is summarized.
      Additionally, the main application scenarios are investigated, including cell networks, vehicular networks, industrial networks, UAV networks, multimedia applications, IoT, blockchain-empowered networks, and energy harvesting-assisted networks.
      Therefore, the learned lessons are summarized to provide some insights.
  \item Finally, the potential directions are highlighted in the RL-empowered MEC.
\end{itemize}

\begin{figure*}[thpb]
\centering
\includegraphics[width=6.9in]{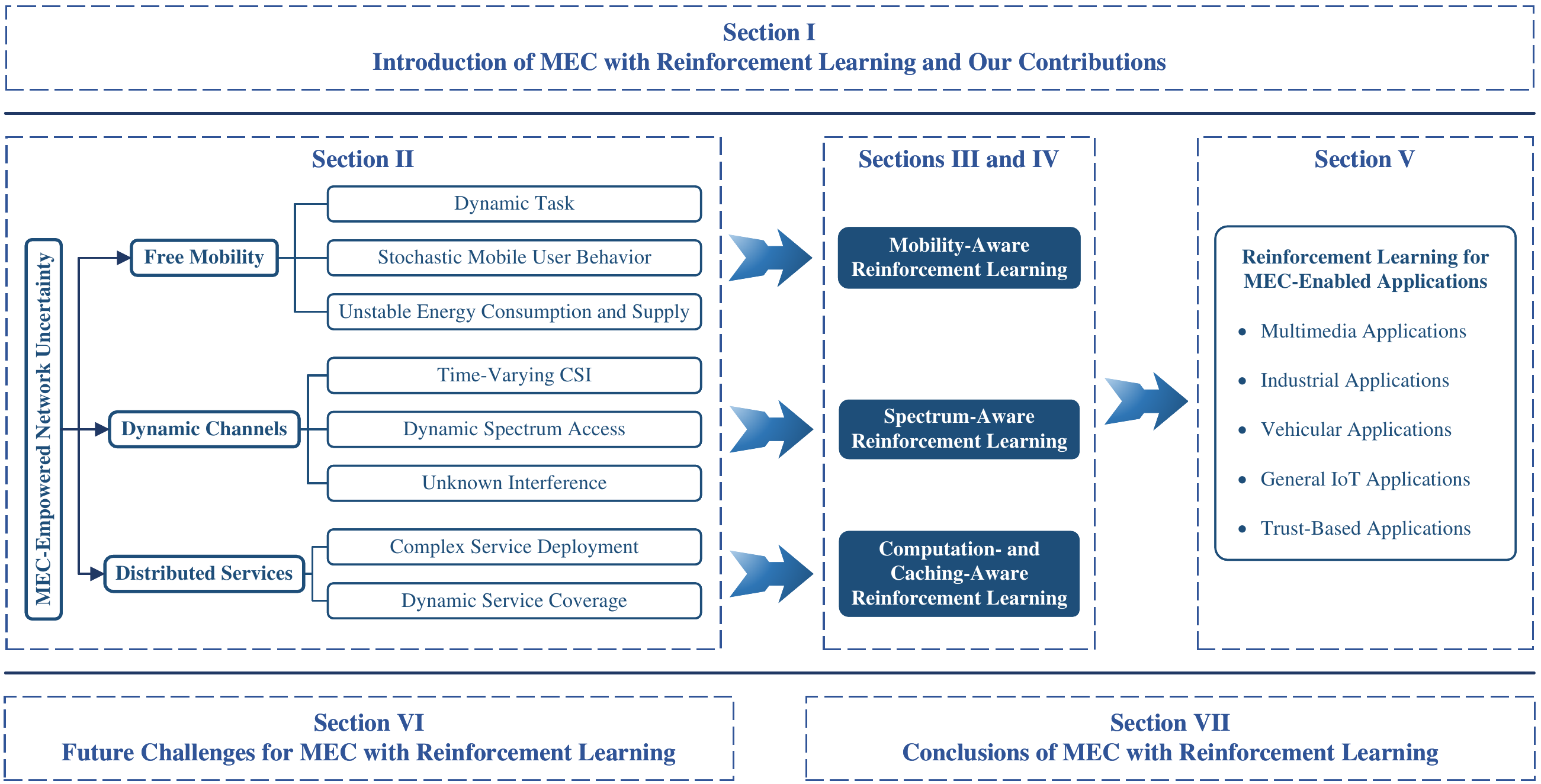}
\caption{Structure of the survey.}
\label{fig:fig4}
\end{figure*}

The remainder of the paper is organized as shown in Fig. \ref{fig:fig4}.
In Section II, we first describe the system model of the MEC and then summarize its uncertainty problems associated with free mobility, dynamic wireless channels, and distributed services.
In Section III, a review of RL is briefly introduced and the RL-based MEC framework including two uncertainty-focused strategies is carefully explained.
In Section IV, state-of-the-art research is reviewed from the perspective of uncertainty problems.
In Section V, diverse applications of multimedia, industry, vehicles, IoT, and trust-based networks are investigated.
In Section VI, we outline the open challenges, followed by the conclusions in Section VII.

\section{System Aspects and Uncertainty Problems}

In this section, we first briefly introduce the MEC system model and its corresponding network optimization.
Thereafter, the uncertainty problems related to the dynamics and randomness are highlighted.

\subsection{System Description}

The decentralized MEC architecture consists of three layers as follows.
{\em At the outermost}, mobile devices such as intelligent vehicles, smartphones/tablets, VR/AR/XR devices, and UAVs are responsible for sensing and collecting raw data from the environment.
With certain computing, communicating, and caching capabilities, each mobile device can \textit{locally} process part of the raw data and send the rest to its corresponding MEC servers for \textit{edge processing}.
{\em In the middle}, in close proximity to mobile devices, MEC servers attached to multiple access points (APs) become a bridge between mobile devices and cloud servers.
These nearby MEC servers can be small data centers, gateways, and switches, which can provide faster services than remote cloud servers.
MEC servers may quickly process the collected raw data and send responses to mobile devices when possible. Otherwise, MEC servers may upload part of the raw data to cloud servers or neighboring MEC servers due to their limited computational capability or heavy network load.
{\em In the center}, cloud servers can provide full services to mobile devices and help MEC servers complete communication, computation, caching, and control tasks.

\begin{figure*}[thpb]
\centering
\includegraphics[width=6.5in]{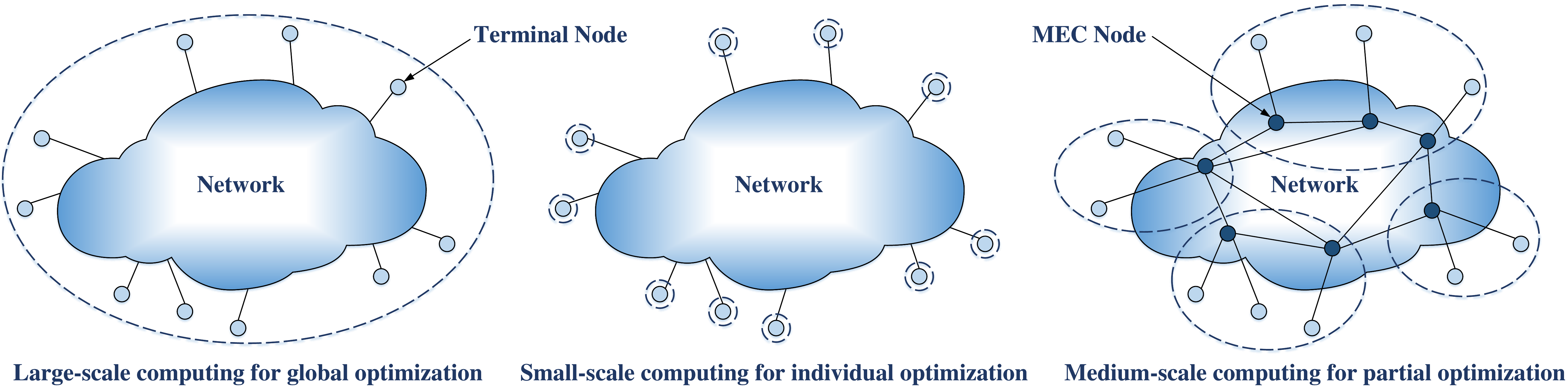}
\caption{Partially decoupled optimization among different MEC-based clusters and partially coupled optimization within each MEC-based cluster.}
\label{newfig:fig2}
\end{figure*}

For the Internet of Everything, heterogeneity is an important feature in MEC systems.
The heterogeneity comes from integrating different types of networks that have different architectures, are equipped with diverse resources, and need to serve various kinds of tasks.
Unfortunately, as shown in Fig. \ref{newfig:fig2}, the individual mobile device is always selfish and does not know the global network state.
As a result, owing to the limited battery life, CPU/graphics processing unit (GPU), and storage, it cannot explore all network environments to make global decisions.
In contrast, the cloud can achieve the global strategy at the expense of considerable training and inference processes.
With the aid of MEC for virtual or physical network infrastructures, decoupled/coarse-grained/simplified network optimization can be achieved in terms of task offloading, resource allocation, task scheduling, etc.
Thus, the hierarchical architecture in MEC is beneficial for using heterogeneous resources to satisfy the heterogeneous requirements in 5G networks and beyond.
However, due to the incomplete observation of the {\it network dynamics and randomness}, named {\it network uncertainty}, the MEC-aided optimization problem solution may be far from the optimal solution and further degrades the QoS or QoE performance for delay-sensitive and computation-intensive applications.
In this regard, we then elaborate on network uncertainty, followed by effective RL-based strategies.

\subsection{Network Uncertainty}

In this subsection, we identify that network uncertainty poses several challenges to successful MEC deployment and service, as it essentially results in many dynamics and randomness.
We now describe the uncertainty problems that may be introduced in three parts of an MEC-based network, namely, at the mobile device side, over the wireless channels, and at the MEC servers.
Correspondingly, a taxonomy of the MEC-based network uncertainty is listed in Fig. \ref{fig:fig4}.

\textbf{First}, due to the free mobility of mobile devices, task generation, arrival, and execution, network topology, and energy supply and consumption are significantly affected.
Specifically, the main representatives of the MEC uncertainty at the mobile device side are summarized as follows.
\begin{itemize}
  \item {\it Dynamic Task}:
  {\it Random task generation and arrival} exist for a mobile device since independent tasks are possibly sensed, collected, and generated by multiple types of applications stochastically.
  They queue in the task buffer if they cannot be processed immediately due to the limited mobile device computation and communication capabilities.
  For delay-sensitive applications, the task buffer queuing state should be characterized in each mobile device to reduce the queuing delay and avoid buffer overflow.
  However, the time-varying task arrival model poses a challenge to the queuing model and further to the computational offloading and resource allocation optimization in MEC.

  \quad Furthermore, due to {\it task dependency and priority}, diverse decisions for inter- and intra-task executions should be made in MEC-empowered networks.
  Since a practical mobile application is composed of multiple components with sophisticated interdependency, the concept of the task call graph is defined to model task dependency \cite{Ref49}, including sequential, parallel, and hybrid dependency \cite{Ref10}.
  Task dependency indicates the causal/logical relationship of different tasks, where the outputs of some tasks are the inputs of others.
  For example, the limited computational resources stimulate the offloading of some tasks from the local to MEC/cloud servers, while other local tasks, such as video playback and image display, start to be processed after receiving the output of these tasks executed at the MEC/cloud servers.
  Additionally, in general networks, tasks may also have different priorities for different applications (task priority).
  For example, for delay-sensitive and mission-critical applications in vehicular networks, depending on the arrival interval or the wireless channel, the tasks are classified and processed \cite{Ref61}.
  Due to the free mobility of mobile devices, a large number of new tasks with unknown priority may appear, and some tasks with known priority may disappear.
  As a result, the task scheduling efficiency is significantly reduced if fine-grained tasks cannot be properly distinguished from excessive categories \cite{Ref62}.
  When task relevance between different mobile devices exists, such as in the wireless sensor network \cite{Ref50}, more complicated relationships are coupled among randomly arrived tasks, thereby raising challenges for efficient task offloading, resource allocation, and task scheduling.

  \item {\it Stochastic Mobile User Behavior}:
  {\it Random user arrival and departure} are in the coverage of an MEC server.
  Specifically, when a current mobile device completes its computational offloading, it enters an inactive state.
  Stochastically, newly arriving mobile devices generate new computation offloading requests.
  Along with the free movement of mobile devices, such stochastic behavior further increases the unpredictability of task arrival in temporal and spatial domains.

  \quad It is also found that in daily life, the distribution density of mobile users has a periodic spatial-temporal aggregation effect, resulting in a {\it time-varying topology} \cite{Ref38}.
  When mobile users enter urban hotspot areas \cite{Ref39}, where a large number of mobile users are accessing the current MEC servers, delay-sensitive mobile applications may not obtain a rapid response, which deteriorates mobile user QoE.
  In addition, mobile device owners always have inseparable social relationships in decentralized opportunistic communication, such as mobility patterns, daily routines, and information preference in mobile social networks \cite{Ref57,Ref58,8647548}.
  Thus, {\it diverse social attributes} are incurred by different mobile user densities and moving patterns in different areas or times.

  \quad It is noted that a hotspot area is always surrounded by some nonhotspot areas.
  In this case, it is challenging for frequent moving users to use the abundant MEC server computing power in nonhotspot areas to dynamically assist MEC servers with insufficient computing power in hotspot areas.
  Due to the finite coverage of each static AP, such as the ground base station (BS), {\it frequent handover} is caused by fast-moving devices, and allowing {\it dynamic user association} may bring benefits in terms of the signal-to-noise ratio (SNR), spectrum efficiency, load balancing, and energy efficiency.
  However, in a large-scale network with time-varying topology, dynamically and effectively associating a large number of mobile devices with various MEC servers is a sophisticated issue.
  Additionally, due to stochastic user arrival and departure, where and when to perform task scheduling become more complicated.

  \item {\it Unstable Energy Consumption and Supply}:
  In practical scenarios, the energy consumption of mobile devices is unstable.
  This is because different mobile applications may consume different energies, and the same application may experience fluctuating energy consumption in a time-varying wireless environment.
  In this regard, the remaining power of the battery-only devices becomes an important indicator for effectively performing the next actions.
  However, the deterministic energy consumption model is difficult to formulate owing to the time-varying states of the remaining power.

  \quad Another energy-critical problem for mobile devices in wireless networks is ensuring sufficient energy supply due to their limited battery life \cite{Ref166}.
  Although the energy consumption performance can be improved by large-capacity batteries and frequent battery recharging, the former increases the hardware size, weight, and cost, and the latter is unfavorable and impossible, such as for mobile phones and hard-to-reach wireless sensor devices \cite{Ref43}.
  Against these issues, a promising wireless powering technology called energy harvesting facilitates self-sustainability, because of its capability of capturing recyclable energy, such as solar radiation, wind, and radio-frequency signals.
  However, violent energy harvesting readily incurs an {\it unstable energy supply}, which may not afford heavy computational offloading through unreliable wireless transmission.
  Task scheduling is also interrupted in wireless powered MEC when energy transfer and data communication appear in the same radio frequency (RF) band.
\end{itemize}

{\textbf{Second}}, the efficient wireless transmission between mobile devices and MEC servers plays a key role in the performance of MEC-based networks.
Based on the dynamic patterns of opportunistic wireless connections, unreliable wireless channels and scarce radio resources may severely degrade the transmission performance.
Here, the main uncertainty factors in the wireless channel are listed below.
\begin{itemize}
  \item {\it Time-Varying CSI}:
  In practice, the channel state information (CSI) between the mobile device and MEC server always varies over time due to the dynamic wireless environment, such as obstacles, weather, and device mobility.
  For example, for an orthogonal frequency division multiplexing (OFDM) signal transmitted at the carrier frequency of 5.86 GHz in dedicated short-range communications (DSRC), when the moving speed increases from 6 mph to 75 mph, the coherence time decreases from 2.5 ms to 0.2 ms \cite{Ref34}.
  Since the execution time of most mobile applications is in the range of tens of milliseconds \cite{Ref35}, the whole execution process of one application may experience many different channel states.
  As a result, due to the time-varying CSI, offloading all tasks to one specific MEC server is not always an advantageous option.
  Instead, opportunistic task offloading between local execution and MEC processing will be an alternation.
  However, to decide the offloaded task size and required radio resources, efficient opportunistic offloading through time-varying wireless channels is still challenging.

  \item {\it Dynamic Spectrum Access}:
  To accommodate a large number of mobile devices and enhance spectrum utilization in MEC-based networks, dynamic spectrum access is a necessary scheme.
  In practice, since mobile devices do not always observe all statuses of wireless channels, they selfishly compete for limited radio spectrum resources to minimize their delay and energy consumption.
  For example, when some specific channels are occupied by Wi-Fi users, severely contaminated signal transmission is used for computational offloading from a long-term evolution over an unlicensed spectrum (LTE-U)-enabled IoT device to the corresponding MEC server, resulting in significant system performance degradation, such as the additional delay caused by data retransmission.
  For a large number of random mobile tasks, the major question is how to orchestrate unlicensed spectrum resources and coordinate time-varying channel access for communication-efficient task execution in the MEC.

  \item {\it Unknown Interference}:
  For a specific mobile device in a cell, its available radio bandwidth may be interfered with by spectrum leakage from adjacent channels, and its signal-to-interference-plus-noise ratio (SINR) is reduced by heavy interference from neighboring cells.
  Consequently, the reduced transmission rate will incur a significant transmission delay for a large quantity of data offloaded to (or downloaded from) an MEC server.
  Furthermore, in practical scenarios, the interference models are always unknown, and can be classified as human-made interference and objective interference.
  Compared to objective interference, human-made interference brings more uncertainties due to the unpredictability of human behavior, such as intensive device-to-device (D2D) communications overlapping with the communication resources used by mobile devices.
  On the one hand, trusted mobile users may unintentionally cause interference, such as cochannel interference from unlicensed users.
  On the other hand, malicious users actively create serious jamming in a stochastic manner.
  Objective interference always comes from a constantly changing environment, such as rain.
  With serious and unknown interference, MEC systems are challenged to make optimal offloading decisions and allocate communication resources.
\end{itemize}

{\textbf{Third}}, the network uncertainties at the MEC server have significant impacts on the optimization, decision, and performance in MEC-based networks.
Specifically, due to the device's mobility and dynamic wireless channel, indeterministic task arrival causes more serious uncertainties in the MEC server.
In detail, the corresponding uncertainty issues are summarized below.
\begin{itemize}
  \item {\it Complex Service Deployment}:
  Compared to centralized cloud service deployment, MEC has more advantages in delay-sensitive and computation-intensive mobile applications due to its decentralized heterogeneous resources proximal to mobile devices.
  With the assistance of network virtualization technologies, such as software-defined networking (SDN) and network function virtualization (NFV) \cite{Ref44}, services can be flexibly deployed in edge servers by decoupling hardware and software.
  However, in addition to throughput, resource constraints, time constraints, response quantity, and energy consumption in the centralized network, more uncertainty factors should be considered in MEC service deployment, including the geographic location and access pattern (e.g., different preferences and time-varying access frequency for different services) of the mobile device \cite{Ref56}.
  Furthermore, the ordered combinations and logical connections of a series of NFV-based components may be involved in the deployed services, such as
  microservices \cite{Ref55} and \textit{service function chains} (SFCs) \cite{Ref51}, which complicate the service relationships.
  Additionally, with the proliferation of smart mobile devices, diverse services are enabled by multiple virtual slices in the physical RAN infrastructure \cite{Ref44,Ref45,Ref46,Ref48}.
  To efficiently deliver multiple services to mobile devices, especially for multimedia services (e.g., immersive VR/AR/XR in video applications), it is challenging to perform multitenant cross-slice resource orchestration in the MEC server to satisfy the distinct QoS requirements of all mobile users.

  \quad One of the most popular MEC service deployment issues is {\it content caching and delivery}.
  Content caching involves static content (such as popular videos) and dynamic content (such as environmental parameters of temperature, humidity, and illuminance).
  The former is typically not changed in a long period for delay-tolerant applications, and in contrast, the latter is updated in a relatively short period for mission-critical or time-sensitive tasks \cite{Ref53}.
  Upon storing the reusable contents in the cache resources at the edge networks, the same content for multiple mobile users can be quickly delivered.
  Additionally, the redundant traffic, repeated calculation, and power consumption in the duplicate transmission are considerably reduced.
  However, in realistic scenarios, mobile users with different habits have diverse task requirements, and many files are uncertainly correlated, such as live footage on televisions and news on websites.
  In addition, different edge devices, including MEC servers and mobile devices, have different and uncertain willingness to share content in terms of content preference, social ties among mobile users, node adjacency, and QoE requirements \cite{Ref54}.
  Consequently, popularity models constantly change over time, in which the critical problems are where, when, and how to efficiently cache appropriate contents in MEC servers.

  \item{\it Dynamic Service Coverage}: Capability-constrained MEC often involves multiple servers to complete challenging tasks, which may require {\it server cooperation}, {\it service migration}, and/or {\it mobile MEC coverage} due to user mobility or service outages.
      We first consider server cooperation and assume that services are statically deployed to each MEC server.
      It is predicted that the number of IoT devices connected to the internet will reach approximately 29 billion in 2022 \cite{Ref37}.
      Thus, a large amount of task offloading will be incurred in IoT applications, which cannot be efficiently processed by a single MEC server, as opposed to the powerful cloud.
      If all tasks were offloaded to the cloud, network congestion may be unacceptable.
      Therefore, the cooperation between MEC and cloud servers and the collaboration between different MEC servers provide beneficial solutions in terms of the complementary components or functions of computing, communication, and caching.
      In this case, the computational tasks are divided into groups and offloaded to multiple MEC servers and the cloud server.
      Different link states appear in multihop communications, which may further increase the transmission delay and spectrum resource consumption.
      Additionally, the resources in MEC servers are frequently occupied and released randomly.
      Therefore, how to make real-time decisions and executions with the appropriate selection of cloud servers or distributed MEC servers becomes a critical challenge for server cooperation.

\begin{figure*}[thpb]
\centering
\includegraphics[width=6.5in]{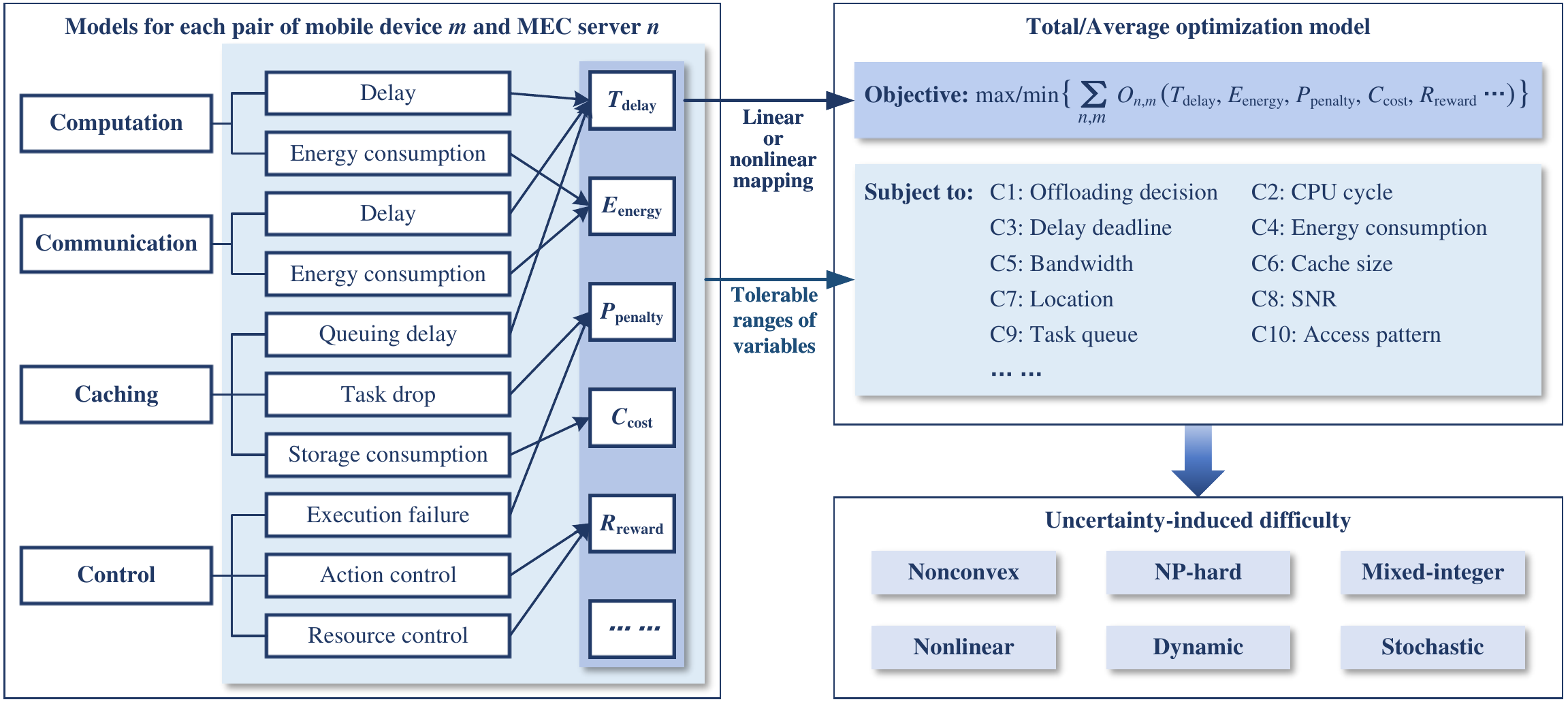}
\caption{Illustration of the MEC optimization problem with multiple coupled models and a large number of different variables.}
\label{fig:fig6}
\end{figure*}

  \quad Service migration is another function that can provide flexible service coverage.
  When the device's mobility is involved in the wireless network with ultradense MEC servers, the service requested by the mobile device from the initial MEC server is now provided by another new MEC server, where the corresponding service provider (e.g., virtual machine (VM) or container) is migrated through one or more hops communications from the initial MEC server to the current counterpart.
  Different switching latencies and energy consumption of the service migration are incurred between different pairs of MEC servers.
  With the explosive growth in mobile devices, reducing operational costs due to frequent service migration is a nontrivial issue.
  Moreover, the time-varying and unpredictable network status in practical systems, such as the network topology and coverages of different APs, makes the problem even more complex.

  \quad Furthermore, due to the static deployment of MEC servers, when services are requested by a tremendous number of mobile users, a temporary service outage may occur for some users in a weak network environment, such as in blind spots.
  To assist the static server in satisfying enormous mobile requirements, movable edge devices can facilitate reliable and cost-effective solutions, such as ground vehicles, UAVs, and satellites \cite{9174846,9183767}.
  Via extending coverage and increasing wireless links, these movable edge devices provide necessary supplements, such as relays and lightweight servers.
  Unfortunately, their high mobility creates more network dynamics, e.g., unstable network topology and unreliable wireless channels, into service provisioning, which further complicates MEC service deployment, such as UAV-based MEC deployment assisted by path planning \cite{9448189,Ref93,Ref96,9354996,Ref95,9397778}.
\end{itemize}

In MEC-based networks, a complete service process is always supported by the operations on the mobile device side (e.g., task request, task offloading, and energy harvesting), the wireless channel side (e.g., access control, interference management, and transmit power control), and the MEC server side (e.g., service deployment, service migration, and caching strategy).
For instance, for delay-sensitive task processing, computational offloading from the mobile device to the MEC server is necessary, which involves local processing, wireless transmission, and MEC server processing.
If the mobile device-, wireless channel-, and MEC-side models are independently considered in the whole system, the offloading execution of the delay-sensitive task may fail due to packet loss with incorrect channel estimation.
Thus, the MEC-based optimization model should consider not only mobile devices, wireless channels, and MEC services but also their coupling relationships.

In summary, as shown in Fig. \ref{fig:fig6}, the network optimization in MEC always involves delay and energy consumption, which is associated with multiple variables, such as offloading and caching decisions, bandwidth provisioning, transmit power control, and CPU frequency adjustment.
These correlated variables indicate that heterogeneous computation, communication, and caching resources are coupled with each other.
For heterogeneous resource management, discrete variables and continuous variables always coexist, such as discrete task offloading variables and continuous resource allocation variables.
In addition, with the enormous number of network devices, a large number of high-dimensional variables will appear.
More importantly, the inherent dynamics and randomness of free mobility, dynamic characteristics of the wireless channel, and distributed MEC services introduced in this section will further complicate the joint optimization problem of the MEC system, such as task offloading and scheduling problems, user association and handover problems, fragmented streaming and resource management problems, path planning and energy saving problems.
Therefore, the optimization problem is always mixed-integer, nonconvex, and NP-hard and is intractable to solve with unknown network states.
To address these challenges, RL-based MEC optimization algorithms are introduced in the next section.

\section{Reinforcement Learning-Based MEC}

In this section, we first briefly introduce RL and then outline the RL-empowered MEC framework using two uncertainty-focused strategies.

\subsection{Reinforcement Learning}

As mentioned, in a production MEC network, we have to deal with uncertainties caused by dynamic/random network characteristics to efficiently collaborate heterogeneous resources and meet heterogeneous QoS and QoE requirements.

To this end, RL is regarded as a promising way through interacting with the dynamic network environment.
It is based on the characterized learning problem and is not restricted to a specific method \cite{Ref116}.
Specifically, in RL \cite{Ref21}, through a proper trade-off between \textit{exploration} and \textit{exploitation}, an agent continuously learns a state-action value function that maps the observed environment states to the most effective actions to maximize a reward.
Different RL techniques can be used, such as dynamic programming (DP), Monte-Calo (MC), and temporal-difference (TD), which involve different types of approximations of the value function and different algorithms to learn the function.

In recent years, deep learning has sparked wide interest, where the state-action value (also known as the {\it Q}-value \cite{Ref181}) function is approximated by neural networks (NNs) of various structures such as DNN, convolutional NN (CNN), and recurrent NN (RNN), etc.
When combined with the {\it Q}-learning algorithm and experience replay, the so-called deep {\it Q}-learning network (DQN) framework can handle fairly complex state space and achieve efficient learning compared with traditional {\it Q}-learning.
In DQN algorithms, since the value-based learning framework suffers from insufficient processing for continuous actions, the policy-based counterpart is designed.
Moreover, by combining the advantages of policy-based and value-based DQNs, an actor-critic network is designed, such as deep deterministic policy gradient (DDPG) \cite{lillicrap2015continuous} and asynchronous advantage actor-critic (A3C) \cite{mnih2016asynchronous}, which are state-of-the-art techniques, as covered when discussing the relevant MEC problems below.

\begin{figure*}[thpb]
\centering
\includegraphics[width=7in]{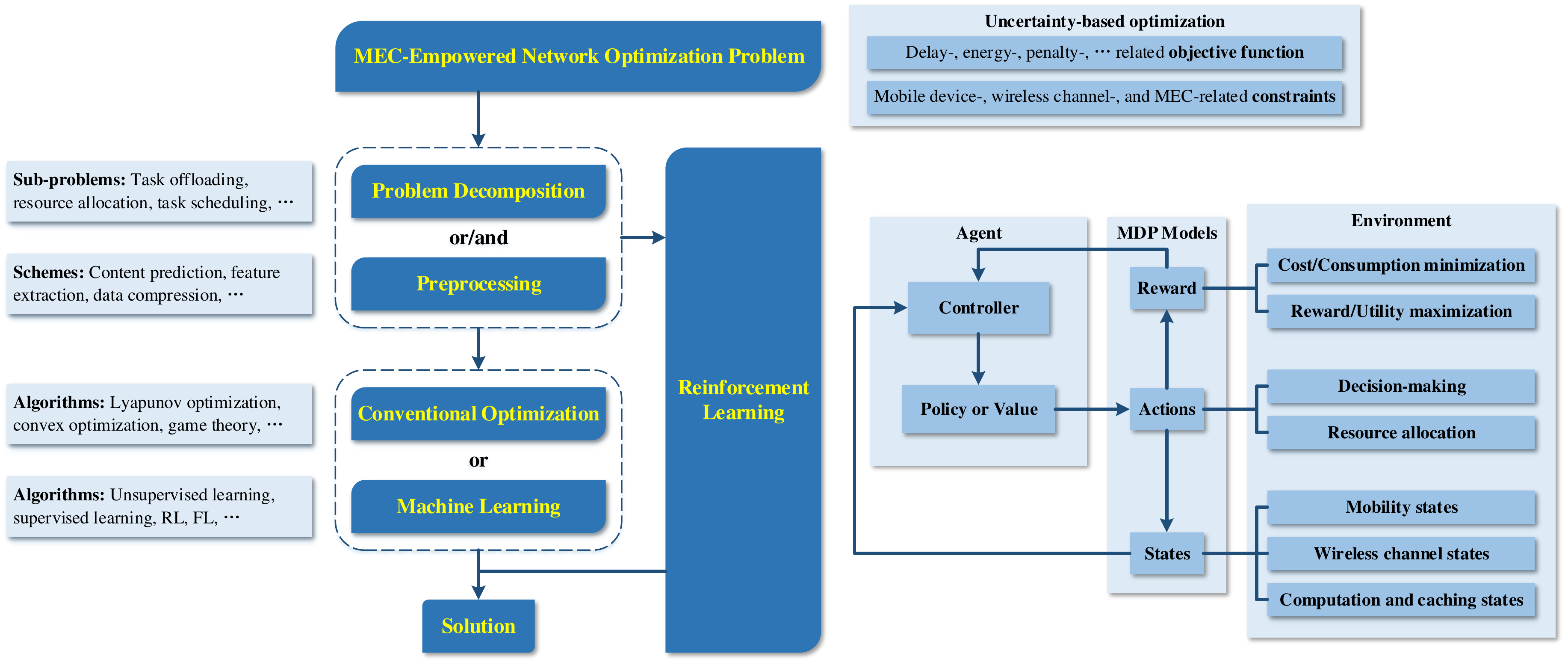}
\caption{A general RL-based optimization framework in MEC systems.}
\label{fig:fig9}
\end{figure*}

\subsection{Reinforcement Learning-Empowered MEC}

Based on the analyzed MEC optimization problem with multiple network uncertainties in Section II and the reviewed RL algorithms in Section III-A, the RL-based MEC optimization framework is summarized as shown in Fig. \ref{fig:fig9}.

Specifically, it is observed in Fig. \ref{fig:fig9} that upon capturing mobility, channel, computation, and caching states in the random environment, the optimal action policy with the maximal reward can be exploited by the trained agent.
For example, after the optimization objective function and its constraints are mapped to the state, action, and reward in the Markov decision process (MDP), the optimal policy of task offloading and resource allocation can be achieved by the RL algorithms, without any a priori knowledge of the network environment dynamics.
Furthermore, the long-term performance of task offloading and resource allocation can be efficiently achieved by using the powerful capability of neural networks and experience replay along with traditional RL algorithms.

According to the decomposability of optimization problems in MEC-based networks, there are two
potential RL solutions.
The first solution (i.e., S1) is to adopt the RL algorithms directly for the original optimization problem or its equivalent transformation.
In the second solution (i.e., S2), the original optimization problem is decomposed first, and then the RL algorithms are employed to solve the subproblems.
More details are given as follows:
\begin{itemize}
  \item[\textbf{S1}.] After mapping the focused objective function and constraints in the MEC-based networks into the states, actions, and reward in the MDP,
      the RL algorithms are directly employed to the original optimization problem, such as in \cite{Ref65,Ref66,Ref77}.
      Among them, many advanced RL algorithms are applied to balance the convergence speed and learning accuracy, such as DDPG and A3C.
      Furthermore, for the input and output of the agent in the RL, some data preprocessing for the state space reduction, such as data normalization in \cite{9380662} and action searching/refinement/classification methods to efficiently reduce the action space \cite{Ref125,Ref123,Ref73,Ref98,Ref87}, may be introduced to enhance the learning efficiency.

  \item[\textbf{S2}.] In this way, the RL algorithms are indirectly utilized for the original optimization problem or its equivalent transformation.
      Specifically, based on primal/dual/hierarchical/partial decomposition \cite{1664999}, the original optimization problem is first decoupled into several subproblems (usually two subproblems, such as the offloading decision subproblem and resource allocation subproblem for computational offloading optimization \cite{Ref122,Ref126}).
      Then, some with large-scale uncertain states and action spaces are solved by RL algorithms with high efficiency, and other subproblems with determined network states are solved by conventional algorithms, such as conventional optimization algorithms \cite{Ref131,9184934,Ref93,Ref142,Ref126,Ref137}, searching algorithms \cite{Ref106,Ref114,Ref122}, heuristic algorithms \cite{Ref80}, and game theory \cite{Ref92,Ref82}.
      Alternatively, integrating RL and machine learning algorithms is adopted, such as supervised/unsupervised learning \cite{Ref59,Ref124,Ref75}, federated learning (FL) \cite{Ref160,Ref74,9426913}, and hierarchical RL \cite{Ref99,Ref76,9397778,Ref148,Ref89}, for optimization efficiency improvement.
\end{itemize}
In large-scale networks, the first solution may not be able to make the best policy of decision-making and resource allocation.
In this case, the second solution identifies a promising direction for computationally efficient network optimization due to the reduced state and action spaces from the problem decomposition.

In addition, to solve MEC system optimization problems, detailed mapping from the objective function and constraints to the MDPs should be determined.
First, the objective function can be directly mapped to the reward function to maximize the system utility (or minimize the system cost), as in \cite{Ref36,Ref73}.
Negative or reciprocal partial or total objective function terms can also be formulated as the reward function \cite{Ref70,Ref74}, which is the most popular expression for the reward function.
In addition, values of $\pm 1$ and $0$ are sometimes adopted for the reward functions, such as in  \cite{Ref69,Ref108}.
Furthermore, partial or total terms of the objective function can also be mapped to some states.
For example, for an objective function with the total delay/energy, the delay/energy consumption state may be observed to evaluate the timeout penalty/service failure penalty.
Second, since the constraints always give the range of all random variables, most constraints are mapped to the states.
In the decomposed subproblems, parts of the constraints are transformed into the objective function, which may be mapped into the reward function.
Action-related constraints are always associated with states and actions.
More importantly, to effectively capture the network uncertainty in MEC systems, corresponding to the original optimization problem involving mobile devices, wireless channels, and MEC servers, the mapped states can be classified as mobility states (e.g., locations of mobile devices, task generation/arrival, and battery level), channel states (e.g., time-varying CSI, wireless bandwidth, and interference), and computation and caching states in the MEC server (e.g., CPU frequency, queuing delay, and content sharing willingness).
Since these states depend on the MEC environment dynamics and randomness, the quantity of coupled states is always large in MEC-empowered networks.
Finally, since the optimization problem solutions are related to decision-making policy and resource allocation, the solutions are always mapped to the actions.
To satisfy the user's requirements in the MEC, the actions mainly involve decision-making of task offloading, task scheduling, service migration, and server selection, and resource allocations of communication, computation, and caching/storage/memory.

\section{State-of-the-Art Research: A Perspective of Uncertain Problems}

In this section, according to the three categories of uncertainty summarized in Section II-B, the MEC empowered network optimizations using the RL solutions are summarized and learned lessons are given in terms of the free mobility, dynamic channel, and distributed service.

\subsection{Studies on Free Mobility}

\subsubsection{Task Arrival and Dependency}

For time-varying task arrival and unknown personal utility information, Ref. \cite{Ref68} adopts policy gradient-based DRL for a dynamic pricing algorithm.
Instead of {\it Q}-learning using the discrete pricing space, a continuous pricing strategy of the offloading service fee is enabled in \cite{Ref68} to maximize the long-term AP revenue.
Simulation results validate the enhanced effectiveness of the proposed dynamic pricing algorithm compared to the conventional {\it Q}-learning algorithm.
Due to the random task arrival in each mobile device, the uncertain load dynamics problem at the MEC servers is addressed in \cite{9253665}.
To minimize the expected long-term cost of task delay and dropping penalty, the optimal offloading decision-making strategy was proposed in \cite{9253665} with the aid of long-short-term memory (LSTM), dueling DQN, and double DQN algorithms.
In \cite{Ref78}, when multiple subtasks of each mobile device are simultaneously uploaded to multiple MEC servers, the time-varying task characteristics and computational capability at the MEC side are unfavorable for conducting highly efficient offloading for real-time applications.
Aiming at minimizing the short-term system cost associated with delay and energy consumption, a dynamic offloading strategy for multiple mobile devices is proposed by DQN-based training and learning.
\cite{Ref68,9253665,Ref78} adopted RL algorithms directly to address computational offloading problems.
In addition, some studies combine the RL algorithm with the traditional optimization algorithm for highly efficient network optimization.
For instance, considering the time-varying channel and random task arrival at mobile devices, binary offloading and resource allocation are modeled as a mixed-integer nonlinear programming (MINLP) problem.
Then, based on primal decomposition, the master problem of offloading decisions is solved by the DNN-based actor network, and the
subordinate 
problem of resource allocation is optimized by the primal-dual algorithm \cite{9449944}.

\begin{table}[!t]
  \centering
  \caption{Notations From Table \ref{tab:tab2} to Table \ref{tab:tab9}}\label{Tab:Notation}
  \begin{tabular}{c|l}
     \hline
  \textbf{Notation} & \multicolumn{1}{c}{\textbf{Definition}} \\ \hline
  \multicolumn{2}{c}{\textbf{Abbreviation and symbol}}\\ \hline
    The. &  Theme abbreviation \\
    Sol. & Potential RL solution \\
     MD  &  Mobile device-related constraints \\
     CH  &  Wireless channel-related constraints \\
     \checkmark  &  Related component in objective function/constraint \\
     -----  &  Unmentioned content in the corresponding reference \\ \hline
  \multicolumn{2}{c}{\textbf{Decision-making}}\\ \hline
     \ding{172}  &  Task offloading  \\
     \ding{173}  & Task scheduling  \\
     \ding{174}  & Service/resource pricing  \\
     \ding{175}  & Server selection \\
     \ding{176}  & Block length selection \\
     \ding{177}  & Video quality level selection \\
     \ding{178}  & Content updating  \\
     \ding{179}  & Path planning for moving/flying direction and distance  \\
     \ding{180}  & Rate configuration/control/selection  \\
     \ding{181}  & Service/Model migration  \\
     {\tiny\textcircled{\tiny{\textbf{11}}}} & Service deployment  \\
     {\tiny\textcircled{\tiny{\textbf{12}}}} & Traffic light control  \\ \hline
  \multicolumn{2}{c}{\textbf{Resource allocation}}\\ \hline
     \ding{182}  & Computational resource allocation  \\
     \ding{183}  & Bandwidth/Spectrum resource allocation \\
     \ding{184}  & Caching/Storage resource allocation \\
     \ding{185}  & Energy/Power control \\ \hline
   \end{tabular}
\end{table}

The task executions are always strongly coupled, during which the task dependency is depicted by a general task graph.
In this regard, \cite{Ref36} optimized computational offloading to minimize the application execution time.
Then, by observing the task queue and battery states in the mobile device and the task queue dynamics in the MEC server, the reward of the application execution time can be maximized using the {\it Q}-learning algorithm in \cite{Ref36}.
It is demonstrated that compared to random offloading and brute-force offloading, {\it Q}-learning with the $\epsilon$-greedy policy can significantly reduce the total execution time while consuming less energy.
In \cite{Ref125}, a low-complexity DRL-based task execution time and energy consumption optimization is proposed to avoid frequent offloading decisions and resource allocation recalculations, considering time-varying wireless channels and dynamic edge computational resources.
Particularly, to enhance the DRL algorithm convergence speed, three improvements are added for the adopted actor-critic network input and output.
In the actor network, the Gaussian noise-added order-preserving quantization scheme is devised to quickly obtain the optimal local action in the high-dimensional action space.
Then, in the critic network, the joint optimization problem is first simplified by the total number of loop-free paths in the task graph, followed by a closed-form expression of the minimal local CPU frequency.
Finally, the one-climb policy-based heuristic method is used to reduce the number of quantized actions, which is input into the critic network for evaluation.
In this way, RL is combined with optimization theory to improve the running time of the proposed algorithm in \cite{Ref125}.
The detailed comparisons of the studies mentioned above are shown in Table \ref{tab:tab2}, where the abbreviation, symbol, and circled number are listed in Table \ref{Tab:Notation}.

\subsubsection{Mobile Device Movement, Arrival, and Departure}
Considering the dynamics of user mobility, task generation, and network environment, \cite{Ref140} achieved the optimal task offloading and resource allocation strategy by an attention-based double DQN algorithm.
Specifically, two online neural networks are adopted to evaluate the action-value functions related to latency and energy rewards.
Furthermore, a context-aware attention mechanism is proposed for the online weight adjustment of each action value.
Considering the stochastic arrival and departure of mobile devices in an infinite time horizon, \cite{Ref137} minimized the system cost, which is the weighted sum of the number and power consumption of active mobile devices, by the joint optimization of task offloading, mobile device scheduling, and power control.
During the decision-making process, many states are involved in task offloading, local computing, and mobile devices' arrivals and departures, including channel gain, offloaded task queue, local task queue, local CPU frequency, task arrival indicator, etc.
The action space is composed of the index and transmit power of the scheduled mobile device and the offloading decision of the newly arrived mobile device.
In this regard, a low-complexity approximated MDP framework is proposed, which includes the analytical value function of a heuristic scheduling policy and the suboptimal scheduling policy with one-step policy iteration.
Since the scheduling-based value function is always unknown in practice, DRL was leveraged in \cite{Ref137} for the baseline scheduling policy.
Based on this, the power control strategy is further provided by a stochastic gradient descent (SGD) algorithm.

\begin{table*}[thpb]
\centering
\caption{Comparisons of RL-Based MEC with Task Arrival and Dependency, and Mobile Device Movement, Arrival, and Departure}
\label{tab:tab2}
\begin{tabular}{|m{0.15in}|c|c|c|m{0.35in}|c|c|c|c|c|m{0.65in}|m{0.65in}|m{0.3in}|m{0.45in}|}
\hline
\multirow{2}{*}{\textbf{The.}} & \multirow{2}{*}{\textbf{Ref.}} & \multicolumn{3}{c|}{\textbf{Objective}} & \multicolumn{3}{c|}{\textbf{Constraints}} & \multirow{2}{*}{\textbf{Sol.}} & \multicolumn{5}{c|}{\textbf{Reinforcement Learning}} \\ \cline{3-8} \cline{10-14}
 &  &
 \textbf{Delay} & \textbf{Energy} & \multicolumn{1}{c|}{\textbf{Other}} & \textbf{MD} & \textbf{CH} & \textbf{MEC} &
 & \textbf{Model} & \multicolumn{1}{c|}{\textbf{Algorithm}} & \multicolumn{1}{c|}{\textbf{Mobility State}} & \multicolumn{1}{c|}{\textbf{Action}} & \multicolumn{1}{c|}{\textbf{Reward}}  \\ \hline
 \multicolumn{1}{|c|}{\multirow{6}{*}{\begin{sideways}Task Arrival and Dependency  \end{sideways}}}  &   \cite{Ref125}  &  \checkmark & \checkmark &\multicolumn{1}{c|}{-----} &  \checkmark & ----- & \checkmark & S1 & ----- &  DRL  & CPU frequency  & \ding{172} &  \multicolumn{1}{c|}{-----}  \\ \cline{2-14}
 &  \cite{Ref36}  &  \checkmark & ----- & \multicolumn{1}{c|}{-----} &  \checkmark & \checkmark & \checkmark &  S1   &    MDP   &   {\it Q}-learning    & Task index, battery level, task queue &  \ding{172}    &  Execution latency  \\ \cline{2-14}
 &    \cite{Ref68}    &  \checkmark & \checkmark & \multicolumn{1}{c|}{-----} & \checkmark & ----- & -----  &  S1    &    MDP  &   Policy-based DRL  &   Task arrival   &  \ding{174} &  Offloading profit    \\ \cline{2-14}
 & \cite{9253665} & \checkmark & ----- & \multicolumn{1}{c|}{-----} & \checkmark & \checkmark & \checkmark & S1 & ----- & Dueling DQN, double DQN & Load level, queue length, queuing delay, task size & \ding{172} & Task delay \\ \cline{2-14}
 &    \cite{Ref78}     &   \checkmark & \checkmark & \multicolumn{1}{c|}{-----}  & \checkmark & ----- & ----- &   S1   &   MDP   &   DRL   &  Task size, offloading rate  &   \ding{172}    &  System cost \\ \cline{2-14}
 & \cite{9449944} & ----- & ----- & Computa-tion rate & \checkmark & \checkmark & \checkmark & S2 & ----- & Lyapunov optimization + actor-critic  DRL + primal dual algorithm & Task queue, energy queue & \ding{172} \ding{182} \ding{183} & Computa-tion rate  \\ \hline
\multicolumn{1}{|c|}{\multirow{2}{*}{\begin{sideways}  Device Movement \end{sideways}}}   & \cite{Ref137} & ----- & \checkmark & Device number & ----- & ----- & ----- & S2  & MDP & RL+SGD & Task offloading queue, local computing, new active device  & \ding{172} \ding{173} \ding{185}  & System cost   \\ \cline{2-14}
    & \cite{Ref140} &  \checkmark & \checkmark & \multicolumn{1}{c|}{-----} &  \checkmark & ----- & \checkmark & S1  & MDP & Double DQN & Remaining energy, data size, required CPU cycle & \ding{172} \ding{182} \ding{185} & Negative delay and energy consumption   \\ \hline
\end{tabular}
\end{table*}

\subsubsection{Renewable Energy}
To handle the high intermittency and unpredictability of the harvested energy in mobile devices, especially for IoT devices with limited battery capacity, RL algorithms provide a promising solution upon interacting with the constantly changing energy status to accumulate experience.
In this way, optimal offloading solutions can be achieved to improve the network performance when incorporating renewable energy into MEC \cite{Ref100,Ref102,Ref103,Ref104,Ref105,Ref106,Ref107,Ref130,Ref164,Ref66}.
Among them, when a hybrid wireless AP with the data and power transmission capabilities for mobile devices is deployed, the scheduling of energy harvesting and task offloading is a critical challenge in MEC \cite{Ref100,Ref107,Ref106,Ref130}.
Ref. \cite{Ref100} exploited high-rate RF communication and low-power backscatter communication for a hybrid offloading strategy, which can strike balanced energy consumption between local computations and task offloading.
Without the self-generated carrier signals, the backscatter communication is in a passive mode by reflecting the incident RF signals through the antenna with specific reflection coefficients.
By observing the local time scheduling and workload allocation and the active and passive offloading states, the proposed DRL agent in \cite{Ref100} can learn the optimal offloading action.
The stochastic mobility of mobile devices will further complicate task scheduling \cite{Ref107}.
Thus, with the stable and convergent control of the soft actor-critic  algorithm, the multiagent DDPG (MADDPG)-based offloading strategy was efficiently conducted \cite{Ref107}.
In addition, in an MEC system with RF-powered mobile devices and a renewable energy-powered MEC server, a renewable energy-aware offloading framework was investigated in \cite{Ref130} to minimize the total energy consumption, bandwidth allocation, task caching, and overflow penalty system cost.
By adding stochastic noise into actions, the optimal solution for task offloading and resource allocation was efficiently explored by the DDPG algorithm in \cite{Ref130}.
More importantly, in \cite{Ref106}, considering time-varying wireless channels, a low-complexity online binary offloading policy was obtained by decomposing the complex optimization problem into two subproblems (i.e., task offloading and resource allocation).
The task offloading subproblem was solved by the DRL algorithm and the resource allocation subproblem was optimized by the one-dimensional bisection search, by which the computational complexity for solving the original optimization problem was reduced significantly.

In addition to the dynamics of wireless charged energy at mobile devices, the wireless channel and task arrival dynamics were further introduced in an ultradense virtualized RAN with multiple BSs \cite{Ref164,Ref66}.
To evaluate the effect of these dynamics on computational offloading, \cite{Ref66} modeled the utility function for a mobile device as the weighted sum of task execution and queuing delays, task drops, the task execution failure penalty, and the payoff of accessing the MEC server.
For such a utility-based optimization problem, conventional RL suffers from the curse of dimensionality owing to the large state space, including the task queue length, the energy queue length, the association state between the mobile device and BS, and the CSI state.
Thus, without the knowledge of the above network dynamics, two online DRL algorithms were developed in \cite{Ref66} with the aid of the double DQN.
In the first algorithm, experience-based double DQN is adopted to make the optimal computational offloading decision.
Then, in the second algorithm, an online deep state-action-reward-state-action (SARSA)-based RL (Deep-SARL) is proposed, where a linear decomposition of the state-action {\it Q}-function is mathematically employed by decoupling the additive structure of the utility function.
Simulation results showed that the second algorithm using multiagent parallel learning outperforms the first algorithm.

\begin{table*}[t]
\centering
\caption{Comparisons of RL-Based MEC with Renewable Energy}
\label{tab:tab10}
\begin{tabular}{|m{0.2in}|c|c|m{0.6in}|c|c|c|m{0.15in}|m{0.25in}|m{0.6in}|m{0.8in}|m{0.25in}|m{0.6in}|}
\hline
 \multirow{2}{*}{\textbf{Ref.}} & \multicolumn{3}{c|}{\textbf{Objective}} & \multicolumn{3}{c|}{\textbf{Constraints}} & \multirow{2}{*}{\textbf{Sol.}} & \multicolumn{5}{c|}{\textbf{Reinforcement Learning}} \\ \cline{2-7} \cline{9-13}
  &  \textbf{Delay} & \textbf{Energy} & \multicolumn{1}{c|}{\textbf{Other}} & \textbf{MD} & \textbf{CH} & \textbf{MEC} &  &  \textbf{Model} & \multicolumn{1}{c|}{\textbf{Algorithm}} & \multicolumn{1}{c|}{\textbf{Mobility State}} & \textbf{Action} & \multicolumn{1}{c|}{\textbf{Reward}}  \\ \hline
    \cite{Ref66}  & \checkmark & ----- &  Penalty, payment & \checkmark & \checkmark & \checkmark &  S1  & MDP   &  Double DQN, Deep-SARL & Task queue, energy queue    &  \ding{172} \ding{185}  &   Long-term utility     \\ \hline
   \cite{Ref106}   &   ----- & ----- & Computation rate  & \checkmark & \checkmark & \checkmark &  S2  &  \multicolumn{1}{c|}{-----}  & DRL + bisection search  &  \multicolumn{1}{c|}{-----} & \ding{172} & Total computation rate  \\ \hline
     \cite{Ref100}  & -----  & \checkmark & \multicolumn{1}{c|}{-----}  & \checkmark & \checkmark & \checkmark  &   S1  &  MDP  & DQN, double DQN  & Energy queue, task queue  & \ding{172} &  Energy efficiency  \\ \hline
     \cite{Ref102}   & \checkmark & ----- & Computation profit   & ----- & ----- & ----- &  S1    &   MDP  &  DQN  & Data type and amount, CPU cycle, battery level & \ding{172} & Computation profits minus delay cost \\ \hline
    \cite{Ref103}   & \checkmark & \checkmark & Data sharing, task drop  & ----- &----- &----- &  S1   & MDP & {\it Q}-learning, DRL  & Battery level, harvested energy  & \ding{172} &  System utility  \\ \hline
    \cite{Ref104}   & \checkmark & \checkmark & \multicolumn{1}{c|}{-----}  & \checkmark &\checkmark &\checkmark  &  S1 &  MDP  &  Actor-critic DRL, MADDPG  &   Data size, battery level, harvested energy, priority   & \ding{172} \ding{182} &  System cost minus penalty  \\ \hline
    \cite{Ref105}   & \checkmark & \checkmark & Privacy level, failure loss, queuing cost &  ----- &----- &----- &  S1   & MDP & {\it Q}-learning &  Data generation, priority, and queue, harvested energy, battery level &  \ding{172}  & System utility \\ \hline
    \cite{Ref107}   &  \checkmark & \checkmark & Cost, failure rate  &  ----- &----- &----- &   S1  & MDP  &   MADDPG  & Remaining data, battery level, harvested energy &  \ding{172} & System reward \\ \hline
    \cite{Ref130}   &   \checkmark &  \checkmark & Bandwidth allocation  &  \checkmark &\checkmark &\checkmark & S1  & \multicolumn{1}{c|}{-----} &  DDPG  & Energy queue, task queue, task offloading  & \ding{183} \ding{185} & Negative total cost   \\ \hline
    \cite{Ref101,Ref163}   &  \checkmark & ----- &  Operation cost & ----- &----- &----- &  S1 & MDP  &  DRL & Task arrival, battery level & \ding{172} & Expected long-term cost   \\ \hline
\end{tabular}
\end{table*}

Moreover, by solving the continuous-discrete action space and coordination problems among mobile devices with energy harvesting, actor-critic DDPG and MADDPG algorithms were, respectively, proposed in \cite{Ref104} to maximize the reward.
Specifically, the reward includes the system cost of normalized execution time and energy consumption and the task drop and task queue overflow penalties.
Regarding renewable energy-powered healthcare IoT devices, with the assistance of the post-decision state of CSI \cite{Ref101}, the privacy-aware task offloading RL algorithm accelerated by a Dyna architecture is adopted to protect the privacy of user location and usage pattern \cite{Ref105}.
In \cite{Ref102}, another low-complexity RL algorithm was developed in energy-harvested MEC systems by introducing an intermediate state.
Specifically, the after-state is defined as the intermediate state after the actions of IoT devices but before the harvested energy arrival, since the data generation and the wireless CSI are independent of the offloading policy.
As a result, the size of the state space is significantly reduced, and an after-state DQN algorithm was developed in \cite{Ref102} to maximize the long-term system reward, including the data computation profits with varying priorities and the execution delay cost.
In \cite{Ref103}, with unknown MEC models (e.g., energy consumption and computation delay), RL- and DRL-based offloading policies were proposed to maximize the system utility, which included the energy consumption, the data sharing gain, the total computational delay, and the task drop loss.
The details of the research on the energy-harvested mobile devices are listed in Table \ref{tab:tab10}.

\subsubsection{Learned Lessons}
As shown in Tables \ref{tab:tab2} and \ref{tab:tab10}, in mobility-aware RL for MEC, the task queue at the mobile device is always observed to capture the task dynamics (generation, arrival, and departure).
The states of mobile device movement, arrival, and departure are indicated by the location of the mobile device or the distance between the mobile device and the AP.
The energy queue/battery level is captured to show the energy state.
Due to the high complexity of conventional {\it Q}-learning, DRL algorithms are widely adopted to make decisions on joint continuous and discrete actions, e.g., joint task offloading and resource allocation.
Owing to the complex inter- and intra-relations of tasks, device mobility, and energy supply, MEC-empowered network optimization that integrates the above three aspects is still an open issue.
For instance, when a self-powered BS with MEC is considered \cite{9351561}, more dynamics are introduced from the MEC side.
As a result, a complicated stochastic optimization problem is formulated, which may not be efficiently solved by direct RL.
In this regard, how to utilize the primal/dual/hierarchical/partial to decompose/simplify the original optimization and achieve an efficient RL-aided strategy is also an open problem.

\subsection{Studies on Dynamic Channel}

\subsubsection{Dynamic CSI}
Considering the time-varying wireless channels in practice, \cite{Ref65} modeled a noncooperative exact potential game and proposed a distributed offloading strategy for multiple mobile users in cellular networks.
To maximize the time-average payoff of each mobile device, i.e., the weighted sum of processed CPU cycles and energy consumption, the model-free {\it Q}-learning algorithm was adopted in \cite{Ref65} by capturing the better response with inertia dynamics.
In this way, the offloading decision and power control are made by each mobile user, without prior knowledge of the wireless CSI.
Based on adaptive learning, costly channel estimations and extra signal overhead can be avoided upon learning the payoffs of all users.
Theoretical analysis and simulation results show that in the specific range of the learning parameter, the proposed multiuser offloading mechanism has an effective convergence of the Nash equilibrium to the expected-payoff game and achieves efficient offloading performance.
Moreover, advanced DRL algorithms, such as DDPG and double DQN, were directly applied in \cite{9380662,9406386,9435782,Ref127} to account for computational offloading problems in time-varying wireless channel conditions.
In \cite{9380662}, in a dynamic MEC environment including time-varying CSI, various task requests, and MEC servers' states, delay-aware and energy-efficient task offloading was investigated with multiple MEC servers.
Particularly, based on normalization-assisted data preprocessing and the clipped surrogate function-based reward representative, the end-to-end DRL algorithm is adopted to make the optimal policy of server selection, task offloading, and resource allocation.
To minimize the end-to-end error probability of transmission and computation \cite{9406386}, the DDPG-based policy of server selection, workload assignment, and time scheduling is proposed in a multiaccess MEC system with one mobile device and multiple MEC servers.
The computation error probability is defined as the probability of the delay violation, that is, the computation time exceeding the required time.
In \cite{9435782}, by introducing a temporal feature extraction network and a rank-based prioritized experience replay, the DDPG algorithm was developed with high convergence speed and stability.
In an MEC system with static MEC servers and UAV- and ground vehicle-enabled mobile MEC servers, owing to the constantly changing channel quality and the number of mobile devices, the joint optimization of online decision-making, resource allocation, mobile MEC node management, and user association is discussed.
Here, two improved RL frameworks are devised, which are DNN-based and DRL-based architectures, to obtain the optimal online solution.
The above DRL algorithm improvements are further applied to the real-time decision-making of heterogeneous MECs in dynamic network environments \cite{Ref127}.

In large-scale MEC systems with a dynamic CSI environment, online offloading and resource allocation decision-making were investigated in \cite{Ref126}.
To reduce the computational complexity, the primal optimization problem is decomposed into the resource allocation subproblem and the offloading subproblem.
Correspondingly, the convex optimization algorithm and DRL algorithm are adopted to address these two subproblems.
In \cite{Ref126}, to rapidly obtain the optimal offloading policy, the DRL algorithm is improved as follows.
The compact state representation is first provided by the regularized stacked autoencoder using unsupervised learning.
Second, the heuristic search algorithm is utilized to find the optimal offloading action.
Finally, the DNN is trained with the aid of the preserved and prioritized experience replay.
The details of the above studies are summarized in Table \ref{tab:tab1}.

\subsubsection{Dynamic Spectrum Access}
Owing to local observations and selfish competition among mobile devices, opportunistic access to unlicensed channels leads to long delays and unreliability of data transmission in the LTE-U-enabled  MEC system.
In this regard, the constrained MDP-based task scheduling strategy was proposed in \cite{Ref139} to maximize the average reward, which was defined as the weighted sum of the delay of computing and transmission and the profit of task completion.
Then, a constrained deep {\it Q}-learning solution is given for task scheduling with the aid of the Lagrange duality.
In detail, a modified version of the original reward function is attained by introducing the Lagrange multiplier weighted computational cost, where the optimal value of the Lagrange multiplier is searched by a subgradient-based algorithm.
To solve the problems of spectrum reuse and communication resource allocation in a D2D-enabled MEC system, policy-based DRL optimization of power control and channel selection was efficiently executed to maximize the system capacity and spectrum efficiency in \cite{Ref108}.
Furthermore, hybrid channel access with nonorthogonal multiple access (NOMA) and orthogonal multiple access (OMA) users was considered in \cite{9373579} to achieve the optimal policy of partial offloading decisions and channel resource allocation using actor-critic DQN.
In the time-varying MEC system with multiple users accessing an individual MEC server, to minimize the total energy consumption, the offloading strategy was proposed in \cite{9462445} by considering the heterogeneous resource requirements and delay constraints in communication and computation.
In this regard, {\it Q}-learning and its low-complexity version, i.e., double DQN, are adopted to achieve the optimal strategy.
In addition, for task offloading from multiple mobile devices to multiple MEC servers through one AP, based on multipointer networks, RL-based neural combinational optimization of minimizing total energy consumption was proposed in \cite{Ref73}.
To achieve feasible offloading action in real-time applications, greedy search with sampling was employed.

In addition to spectrum access for computation offloading from mobile devices to MEC servers, big data transmission from MEC servers to mobile devices was also considered in \cite{Ref110}, such as video compression and transmission for movie watching.
In \cite{Ref110}, three different types of computational tasks were requested by multiple mobile devices, which caused dynamic and complex computational task scheduling in the downlink.
Therefore, to minimize the maximal computation and transmission delay among all users, a multistack RL-based resource allocation algorithm is proposed.
The convergence rate of the proposed algorithm can be further enhanced by using the historical resource allocation and user information recorded in the BS stacks.
Furthermore, to support more diverse services over a general network infrastructure, the concept of RAN-only slicing was adopted in MEC-based cellular networks \cite{Ref111}.
In this concept, different air interface parameters are configured to RAN slices with different communication and computational capabilities to satisfy the heterogeneous requests of mobile devices of multiple service providers.
Since each service provider selfishly competes for limited wireless channel resources to maximize its expected long-term payoff, competitive multitenant cross-slice resource orchestration is formulated as a stochastic game under the multiagent MDP for multiple service providers, considering global network dynamics.
The payoff is defined as the difference between the priced utility and the payment, where the utility is associated with the packet queuing delay, the packet drops, and computation and transmission energy consumption.
Then, the SDN-enabled orchestrator is enabled to allocate global wireless communication resources according to the bids of multiple service providers.
To achieve the Nash equilibrium, an abstract stochastic game is constructed with the conjectures of channel auctions among all service providers.
Finally, through the linear decomposition for the state-value function of the MDP per service provider, the low-complexity double DQN-based online slicing policy is obtained.
These involved studies are summarized and compared in Table \ref{tab:tab1}.

\subsubsection{Unknown Interference}

Due to the unknown interference/jamming, edge computing model, and task generation/arrival model, a high time and energy consumption of task offloading may be caused in practice.
In \cite{Ref71}, an RL-based mobile offloading policy was proposed for caching-aided MEC to resist the interference caused by trusted devices and malicious devices.
Computation offloading is executed according to the content popularity, and thus, the optimization objective includes not only the computational delay and energy consumption of mobile devices but also the received SINR and task sharing gain at the MEC server.
The task sharing gain is evaluated by the future reduced computational overhead at the MEC server.
Since the offloading strategy depends on the risk level and {\it Q}-values, the action space involves the available MEC servers, the feasible offloading rate, and the effective transmit power level.
In the state space, the content popularity and the wireless bandwidth in the current time, as well as the energy consumption, the received SINR of the offloading signals, and the computational delay in the previous time, should be considered.
To handle the high-dimensional state space and mixed continuous-discrete action space, four deep neural networks composed of two actor networks and two critic networks are designed for efficient offloading policy learning.
Simulation results show that the proposed scheme can significantly reduce the computational latency while saving large energy under the condition of uncertain jamming and interference.
Owing to spectrum reuse-induced intercell interference, as well as diverse requirements and dynamic wireless environments in small cell networks, the joint optimization of task offloading and interference coordination was proposed in \cite{9485089}.
To minimize the total energy consumption with specific task delay requirements, a multiagent DRL strategy for task offloading and resource allocation is proposed.
Based on the refined strategies for independent actions of small cell BS agents, intercell interference is coordinated by collaborative exploration in the multi-BS environment.
To reduce the signal overhead and computational complexity in collaborative exploration, the original optimization problem is decomposed into multiple single small cell-based subproblems.
By combining DRL for each subproblem optimization with FL for multiagent collaboration, the low-overhead strategy is efficiently attained.

\subsubsection{Learned Lessons}
From Table \ref{tab:tab1}, the channel gain, bandwidth, and interference are always captured to calculate the SNR or transmission rate in wireless communications.
On this basis, communication-efficient offloading and resource allocation can be attained.
Since considerable state and action information should be exchanged between mobile devices, MEC servers, mobile devices and MEC servers, direct and independent RL algorithms may result in low communication-efficient training and inference through the spectrum-constrained wireless channel.
Thus, the integration of RL and other methods (e.g., convex optimization, FL) is an alternative strategy to support the requirements of multiple mobile devices.
In this way, lightweight communication can be achieved by optimization/learning methods, such as FL.
There are still many open problems where the uncertainty derived from the dynamic channel should be addressed. For instance, in large-scale networks, each MEC server observes only limited-coverage channel states, and it remains open to efficiently and globally allocate wireless resources to avoid large and frequent information interactions among servers.

\begin{table*}[hptb]
\centering
\caption{Comparisons of RL-Based MEC with Dynamic CSI, Spectrum Access, and Unknown Interference}
\label{tab:tab1}
\begin{tabular}{|m{0.15in}|c|c|c|m{0.41in}|c|c|c|c|m{0.3in}|m{0.6in}|m{0.55in}|m{0.3in}|m{0.6in}|}
\hline
\multirow{2}{*}{\textbf{The.}} & \multirow{2}{*}{\textbf{Ref.}} & \multicolumn{3}{c|}{\textbf{Objective}} & \multicolumn{3}{c|}{\textbf{Constraints}} & \multirow{2}{*}{\textbf{Sol.}} & \multicolumn{5}{c|}{\textbf{Reinforcement Learning}} \\ \cline{3-8} \cline{10-14}
 &  &
 \textbf{Delay} & \textbf{Energy} & \multicolumn{1}{c|}{\textbf{Other}} & \textbf{MD} & \textbf{CH} & \textbf{MEC} &
 &
 \textbf{Model} & \multicolumn{1}{c|}{\textbf{Algorithm}} & \textbf{Channel State} & \textbf{Action} & \multicolumn{1}{c|}{\textbf{Reward}}  \\ \hline
\multicolumn{1}{|c|}{\multirow{6}{*}{\begin{sideways} Dynamic CSI \end{sideways}} }    & \cite{Ref65} & ----- & \checkmark &  CPU cycle &  \checkmark & ----- & -----  &      S1      &   Game   &  {\it Q}-learning  & Transmission rate    &  \ding{172} \ding{185}  &  Expected payoff  \\ \cline{2-14}
      &   \cite{Ref126}  & \checkmark & ----- & \multicolumn{1}{c|}{-----}  &   \checkmark & \checkmark & \checkmark  &  S2 & \multicolumn{1}{c|}{-----} &  DRL+convex optimization  & CSI  & \ding{172} &  Sum of task delay  \\ \cline{2-14}
  & \cite{9380662} & ----- & \checkmark & \multicolumn{1}{c|}{-----} & \checkmark & \checkmark & \checkmark & S1 & MDP & Double DQN &  Transmission rate & \ding{172} \ding{182} & Negative delay and energy consumption \\ \cline{2-14}
     & \cite{9406386} & ----- & ----- &  Error probability & \checkmark & \checkmark & \checkmark & S1 & MDP & DDPG & SNR  & \ding{175}  \ding{176}  \ding{182} & Negative logarithm of error probability \\ \cline{2-14}
  &  \cite{9435782} & \checkmark & \checkmark & \multicolumn{1}{c|}{-----} & \checkmark & \checkmark & \checkmark & S1 & \multicolumn{1}{c|}{-----} & DDPG & CSI & \ding{172} \ding{182} \ding{183} & Negative cost  \\ \cline{2-14}
    &  \cite{Ref127}   &  \checkmark & ----- & \multicolumn{1}{c|}{-----}   & \checkmark & \checkmark & \checkmark &  S2 & \multicolumn{1}{c|}{-----} &  Clustering algorithm+ DNN+DDPG  & CSI &  \ding{172} & Reciprocal of total delay  \\ \hline
\multicolumn{1}{|c|}{\multirow{7}{*}{\begin{sideways} Dynamic Spectrum Access \end{sideways}}}
  &   \cite{Ref73}    &  ----- & \checkmark & \multicolumn{1}{c|}{-----}  & \checkmark & \checkmark & \checkmark &   S1   &  \multicolumn{1}{c|}{-----}  &   DRL+greedy searching   &   Transmission rate    & \ding{172} \ding{185}  & Energy consumption \\ \cline{2-14}
    &  \cite{Ref108}   &   ----- & ----- &  System capacity  &  \checkmark & \checkmark & -----  &  S1 & MDP & Policy-based DRL  & Communica-tion mode, CSI  & \ding{183} \ding{185}  &  D2D and cellular transmission reward \\ \cline{2-14} 
& \cite{Ref139} & \checkmark & ----- & Complet-ion profit & \checkmark &\checkmark &\checkmark & S1 & MDP & DQN &  Phase counter and indicator, backoff stage & \ding{172}  & Sum of positive profit and negative delay  \\ \cline{2-14}
  & \cite{9373579} & \checkmark & \checkmark & \multicolumn{1}{c|}{-----} & \checkmark &\checkmark &\checkmark & S1 & \multicolumn{1}{c|}{-----} & Actor-critic  DQN & CSI &  \ding{172}  \ding{183} & Negative overhead  \\ \cline{2-14}
  & \cite{9462445} & ----- & \checkmark & \multicolumn{1}{c|}{-----} & \checkmark &\checkmark &\checkmark & S1 & MDP & {\it Q}-learning, double DQN &   Spectrum resource & \ding{172} \ding{182} \ding{183} & Negative energy consumption \\ \cline{2-14}
     &  \cite{Ref110}   & \checkmark & ----- & \multicolumn{1}{c|}{-----} & \checkmark &\checkmark &\checkmark &  S1 & \multicolumn{1}{c|}{-----} & Multistack RL & Transmission delay and resource & \ding{183} \ding{185} & Delay difference \\ \cline{2-14}
    &  \cite{Ref111}   & \checkmark & \checkmark & Packet drops, payment & ----- &----- &----- & S1  & MDP &  Double DQN  & Location for CSI  & \ding{172} \ding{173} \ding{183}  &  Expected long-term payoff  \\ \hline
\multicolumn{1}{|c|}{\multirow{2}{*}{\begin{sideways}  Interference \end{sideways}}}  & \cite{Ref71}   &    \checkmark & \checkmark & Sharing gain, SINR  & \checkmark &\checkmark &\checkmark  &  S1    &   MDP   &  {\it Q}-learning, actor-critic  DRL   &   Bandwidth, SINR  &   \ding{172} \ding{175} \ding{185} &  Offloading utility  \\ \cline{2-14}
   & \cite{9485089}   &   ----- & \checkmark & \multicolumn{1}{c|}{-----} & \checkmark &\checkmark &\checkmark &  S2    &  Multi-agent MDP   & Multiagent DRL + FL   & CSI, interference power & \ding{172} \ding{182} \ding{183} \ding{185} & Negative energy consumption and delay \\ \hline
\end{tabular}
\end{table*}

\subsection{Studies on Distributed Service}

\subsubsection{Complex Service Deployment}
In MEC, service deployment at distributed servers is complicated by dynamics from free mobility and wireless channels.
In \cite{Ref143,Ref55,Ref56,9380677}, with the aid of virtualized technology (e.g., SDN and NFV), service deployment was discussed, as shown in Table \ref{tab:tab4}.
In \cite{Ref143}, for an MEC system with multiple virtual network function (VNF) components, the SFC scheduling problem was formulated as the job-shop scheduling problem.
To complete a specific mobile device task, where to place the VNF and how to transmit a series of network function instances between microdata centers complicate conventional optimization.
Therefore, the DQN-based RL algorithm is adopted to attain the optimal scheduling solution, while the average resource utilization is maintained.
In \cite{Ref55}, upon considering vehicle mobility, the microservice deployment problem was investigated, where the same microservice was deployed on a part of MEC servers.
To minimize the overall service delay with the minimal migration cost, the MDP-based microservice deployment problem is first formulated and then solved by an RL-based online algorithm to obtain a significant solution close to the optimal performance.
To effectively deploy services into resource-constrained MEC servers while satisfying massive low-latency requirements, \cite{Ref56} formulated an optimization problem with the objective as the number of requests per second on MEC servers and the constraints as data dependency, diverse request patterns, and limited edge resources.
The request pattern implies the access pattern of enormous requests in MEC servers.
Owing to diverse access patterns, it is difficult to accurately describe the request patterns with a specific mathematical model.
Thus, the dueling DQN algorithm is employed in \cite{Ref56} to guide service deployment.
In \cite{9380677}, the DRL-based joint optimization of resource provisioning and service placement was developed, attributed to dynamic demands from multiple locations, limited resources at the MEC servers, and increased number and complexity of services.

Another important service deployment problem is content deployment in MEC, which should be addressed considering dynamic content popularity.
First, we discuss the content caching issue with task offloading.
In \cite{Ref124}, the task popularity prediction problem was first solved by an LSTM algorithm based on real-time recurrent learning.
Based on this, similar to \cite{Ref169}, the joint optimization problem of task offloading, resource allocation, and content caching is formulated for total energy consumption minimization.
The formulated problem is a long-term stochastic optimization problem with multiple dynamics, e.g., channel gain coefficients, and thus, the single-agent {\it Q}-learning-based resource allocation algorithm is proposed.
Furthermore, to avoid the large-scale state space and action space, the multiagent {\it Q}-learning algorithm is employed for low-complexity task offloading.
To reduce the randomness of the $\epsilon$-greedy exploration, Bayesian learning automata are introduced for the optimal action selection.
Extensive simulation results show that the advanced RL algorithms significantly outperform conventional RL algorithms.
In \cite{Ref75}, the gated recurrent unit (GRU) algorithm was adopted to predict the task popularity to determine the computation results cached in MEC servers.
Then, with the aid of cooperative caching among servers, aiming at minimizing the total task delay, multiagent DQN-based RL is employed to jointly optimize caching and computational offloading.
In \cite{Ref74}, to handle the popular content replacement in the decentralized MEC system, an intelligent framework that integrates DRL and FL is proposed.
In the integrated framework, with the defined content popularity, distributed DRL is enabled for computational offloading.
More importantly, to deal with DRL agent deployment issues, including limited communication resources, unbalanced and non-independent and identically distributed (i.i.d.) data, and privacy issues, the FL is introduced.
By training the agents in mobile devices, MEC servers, and cloud servers, the proposed integration learning in \cite{Ref74} can achieve near-optimal performance with low overhead, as opposed to conventional centralized double DQN.
In \cite{Ref54}, D2D cache-aided offloading was introduced by sharing the cached content with neighboring mobile devices.
In a distributed caching system using neighboring mobile devices and MEC servers, a blockchain-based caching and delivery market mechanism is proposed.
Correspondingly, to solve the caching placement subproblem, the DQN algorithm is adopted by observing the sharing willingness, cache amount, and content delivery reward.
Based on this, another subproblem of smart contract execution node selection is also solved by the DQN algorithm by observing the current caching placement, computational resource, wireless bandwidth, and trustworthy features.

Then, the content caching issue without task offloading is introduced below.
In \cite{Ref148}, considering the tidal effects in an MEC-enabled network with multiple mobile users and multicast data, a joint pushing and caching strategy was proposed by predicting future mobile users' requests.
Based on infinite-horizon MDP, the optimization problem is formulated as the maximization of the time-averaged transmission cost from the core network to BS and from BS to mobile users.
According to the transmission independence among the core network, BS, and mobile user, the authors decompose the joint optimization problem into the mobile user caching subproblem and the BS caching subproblem.
Based on the long-term content popularity and short-term temporal correlations of mobile user requests, the {\it Q}-learning algorithm is employed for mobile user caching by approximating the value function and decoupling users and actions.
Additionally, the double DQN is adopted for the BS caching.
In \cite{Ref53}, the low-complexity suboptimal solution for the queue-aware cached content update is achieved by the enforced decomposition and the DDPG algorithm to minimize the average age of information (AoI) of the time-varying contents.
When multiple mobile vehicles access the popular contents stored in the MEC servers of the local area data network, the limited resources at the fronthaul links and the discrimination of QoS requirements of different vehicles are two critical issues in downlink radio access.
Specifically, to address the resource starvation problem, based on the formulated Lyapunov function, a stochastic optimization of maximizing the number of connected BSs is first proposed in \cite{Ref142}.
Then, under the optimal resource allocation, to satisfy different QoS requirements, the optimal decision between feedback- and feedbackless-based downlink transmissions is achieved by using the RL-based multiarmed bandit (MAB) algorithm.
Detailed comparisons of studies on caching-aided MEC using RL are also shown in Table \ref{tab:tab4}.

\begin{table*}[!t]
\centering
\caption{Comparisons of RL-Based MEC with Service Deployment and Content Caching}
\label{tab:tab4}
\begin{tabular}{|m{0.15in}|c|c|c|m{0.4in}|c|c|c|m{0.12in}|m{0.25in}|m{0.6in}|m{0.6in}|m{0.3in}|m{0.6in}|}
\hline
\multirow{2}{*}{\textbf{The.}} & \multirow{2}{*}{\textbf{Ref.}} & \multicolumn{3}{c|}{\textbf{Objective}} & \multicolumn{3}{c|}{\textbf{Constraints}} & \multirow{2}{*}{\textbf{Sol.}} & \multicolumn{5}{c|}{\textbf{Reinforcement Learning}} \\ \cline{3-8} \cline{10-14}
 &  &
 \textbf{Delay} & \textbf{Energy} & \multicolumn{1}{c|}{\textbf{Other}} & \textbf{MD} & \textbf{CH} & \textbf{MEC} &
 &
 \textbf{Model} & \multicolumn{1}{c|}{\textbf{Algorithm}} & \multicolumn{1}{c|}{\textbf{MEC State}} & \multicolumn{1}{c|}{\textbf{Action}} & \multicolumn{1}{c|}{\textbf{Reward}}  \\ \hline
\multicolumn{1}{|c|}{ \multirow{4}{*}{\begin{sideways} Service Deployment \end{sideways}} }   &  \cite{Ref56}    &  ----- & ----- & Request number &  \checkmark & \checkmark & \checkmark & S1  & MDP & Dueling DQN   & Possible service deployment  & {\tiny\textcircled{\tiny{\textbf{11}}}}  & Request number difference  \\ \cline{2-14}
   &   \cite{Ref55}   &  \checkmark & ----- & Migration cost   & \checkmark & \checkmark & \checkmark  &  S1 & MDP & {\it Q}-learning   & Available MEC server, instance configuration & \ding{175} &  Negative utility  \\ \cline{2-14}
   &   \cite{Ref143}   & \checkmark & ----- & \multicolumn{1}{c|}{-----} &  \checkmark & \checkmark & \checkmark  & S1  & \multicolumn{1}{c|}{-----} &  DQN  & Residual processing time & \ding{173} & Sum of processing reward, VM reward, and load balance   \\ \cline{2-14}
  & \cite{9380677} & \multicolumn{3}{m{1.4in}|}{Cost of application load, resource overload, container priority, service distance} & ----- & ----- & \checkmark  & S1 & MDP & DQN & CPU, memory, current service and agent & \ding{182} & Cost \\  \hline
\multicolumn{1}{|c|}{\multirow{7}{*}{\begin{sideways} Content Popularity and Caching  \end{sideways}}} &   \cite{Ref53}   &  ----- & ----- &  AoI   &   \checkmark & ----- & \checkmark &  S2 & MDP &  Enforced decomposition +root-finding algorithm +DDPG  & AoI, request number, request queue  & \ding{178} &  Average cost   \\ \cline{2-14}
  &   \multirow{2}{*}{\cite{Ref54}}   & \multirow{2}{*}{-----} & \multirow{2}{*}{-----} & \multirow{2}{0.4in}{Sharing and caching rewards}  &  \multirow{2}{*}{-----} &  \multirow{2}{*}{-----} &  \multirow{2}{*}{-----} & \multirow{2}{*}{S2}  & \multirow{2}{*}{MDP} & \multirow{2}{*}{DQN + DQN} & Delivery reward, sharing willingness, cache amount  &  \ding{184}  & Traffic offloaded to local caching   \\ \cline{12-14}
    &  & &   &    &     &    &   &     &  &  & Computation and caching resources, trustworthy feature  & \ding{175} & Reciprocal of consensus delay    \\ \cline{2-14}
&  \cite{Ref124}   & ----- & \checkmark & \multicolumn{1}{c|}{-----}  & \checkmark& \checkmark & \checkmark & S2  & \multicolumn{1}{c|}{-----}  &  LSTM learning+ single-agent /multiagent {\it Q}-learning +Bayesian learning & \multirow{2}{0.6in}{Energy consumption}  & \ding{172}\ding{182}\ding{184} / \ding{172} & Energy consumption difference \\ \cline{2-14}
     &  \cite{Ref142}  &  ----- & ----- &  Connected BS number  &  \checkmark & ----- & \checkmark  & S2  & MAB & Stochastic optimization + RL  & \multicolumn{1}{c|}{-----} & \ding{183}  & Cost of data drop, transmission delay, energy consumption    \\ \cline{2-14}
     & \cite{Ref75}  &  \checkmark & ----- & \multicolumn{1}{c|}{-----}  &  \checkmark & \checkmark & \checkmark  &  S2   &  \multicolumn{1}{c|}{-----}  &  GRU prediction + multiagent DQN  & Computation delay, available CPU frequency &  \ding{172} \ding{184}  &   Local delay minus total delay   \\ \cline{2-14}
     & \cite{Ref74}    &  \checkmark & \checkmark & Task drop and failure & \checkmark  & \checkmark & \checkmark   &   S2   &   MDP  &  Double DQN + FL &  Energy consumption &   \ding{172} \ding{185} & QoE \\ \cline{2-14} 
     &  \cite{Ref148}  & ----- & ----- & Bandwidth cost    &  \checkmark & \checkmark & \checkmark & S2 & MDP & {\it Q}-learning + DQN  &  Cache state  & \ding{178} \ding{183}  &  Transmission bandwidth cost    \\ \hline
\end{tabular}
\end{table*}

\subsubsection{Mobile MEC Server}

To satisfy the extensive service requests of a tremendous number of mobile devices, vehicle- and UAV-aided network architectures have been proposed with mobile MEC servers \cite{Ref91,Ref92,Ref93,Ref94,Ref95,Ref96,Ref97,Ref98,Ref99,9421372,9336659,9366889,9354996,9397778,Ref134,9426913}.
Due to the flexible coverage of the movable MEC servers, the computational service range is sufficiently extended.
However, the network uncertainty in the MEC environment is further increased by mobile MEC servers.
In \cite{Ref91}, both vehicular and static MEC servers were considered to provide computation services for mobile devices.
Owing to the uncertainty caused by dynamic vehicle traffic, stochastic task requests, and time-varying wireless channels, the utility maximization problem is modeled as an MDP problem, and then, a vehicle-assisted offloading and resource allocation scheme is presented by {\it Q}-learning and DRL algorithms.
Numerical results demonstrate that the proposed scheme can achieve higher utility and lower task delay than conventional counterparts.
On city roads with MEC-enabled traffic lights, for large-scale vehicular networks, static MEC servers, vehicular MEC servers, and static and vehicular MEC servers are inevitable for handling the increased packet collisions and drops, as well as the increased overhead in the medium access control and physical layers.
In this regard, \cite{9421372} suggests dynamic collaborative orchestration among different MEC servers and uses the road pheromone for multiagent DRL to efficiently control the traffic light, which can minimize the average waiting time of all vehicles and reduce the urban traffic congestion.
Moreover, due to security risk, privacy violation and safety hazard activities conducted by malicious service providers, a lack of trust between vehicles further increases offloading uncertainty.
For safe content offloading to vehicular MEC and static MEC, a hierarchical blockchain framework with the characterized trust management mechanism is designed in the vehicular cloud network \cite{9336659}.
Particularly, by capturing the heterogeneous resource distribution and distinctive requirements per blockchain, a multiagent DRL with centralized training and decentralized execution is proposed for smart computation offloading.

In UAV-aided MEC, UAV-enabled service provisioning for mobile users will become complex \cite{9448189} due to the limited capabilities of UAV servers, as well as the constantly changing user demand, geometrical distance, and obstacle risk.
Since UAV-aided MEC servers may be privately owned by different service providers, they have no cooperation with each other and selfishly take execution to maximize their own rewards.
As a result, game theory is a powerful tool for modeling the resource sharing of different MEC servers. However, conventional games cannot be directly adopted due to the partial observation of all MEC servers and unknown network dynamics.
For this reason, \cite{Ref92} designed a two-level stochastic coalitional game model, where a cooperative game is in the upper level and many noncooperative subgames are in the lower level.
In the upper level, the coalition selection problem is first solved by stochastic coalition formation.
Then, in the lower level, for the selected coalition, the RL algorithms are employed to make the optimal offloading decision.
It is proven in \cite{Ref92} that the joint coalition formation and RL algorithm can converge to strongly stable states with the mixed-strategy Nash equilibrium among multiple UAVs.
Then, in the IoT networks with UAV-enabled MEC and blockchain-as-a-service, in addition to the partial observation during the interactions between terrestrial and UAV-enabled MEC servers (leaders), similar to \cite{Ref92}, the blockchain peers (followers) in \cite{Ref99} also have incomplete offloading decision information of other peers when they collect IoT device data, which is uploaded to UAVs.
Therefore, with random environmental states and unknown transition probabilities, a stochastic Stackelberg game is first modeled to capture the interactions between leaders and followers.
Then, based on the MDP/partially observable MDP (POMDP) models, a hierarchical RL algorithm for decision-making and resource allocation is proposed.
In the upper level, the optimal strategy of service cost assignments (e.g., forwarding tasks from peers to the cloud) and flight parameters for UAVs is estimated by RL or the low-complexity DRL.
At the lower level, the optimal offloading decision among peers is achieved by Bayesian RL or Bayesian DL.
Considering the limited battery capacity and computational resources of UAVs, the cloud-edge computing framework was designed in the multi-UAV-enabled MEC system in \cite{Ref97}.
Due to diverse server selection, a mixed-integer task offloading programming problem from multiple mobile devices is incurred.
To minimize the total time and energy consumption of all mobile devices, {\it Q}-learning-based task offloading and resource allocation were proposed in \cite{Ref97}.
Furthermore, considering the renewable power supply for a UAV with stochastic task arrival through the time-varying channel, a model-free DRL-based computation offloading policy was proposed in \cite{Ref98}, where $K$-means algorithm-based classification was adopted to reduce the action space.
In \cite{9366889}, UAV cooperation was discussed for emergency situations in multi-UAV networks, where DDPG-based joint optimization of offloading policy and resource allocation was proposed with the aid of clustered UAVs.
With the aid of the centralized control of a ground BS, a centralized DRL framework was proposed in \cite{Ref134} for UAV cooperation.
Specifically, the agent at the BS is rapidly trained by the double DQN using state representation learning, and then the optimal action of resource allocation is executed by UAVs using the trained information.
Furthermore, considering the communication failure between the BS and UAVs, distributed DRL-based UAV cooperation for offloading and resource management was developed in \cite{Ref134}, where UAVs transfer the optimal {\it Q}-value without the control of the BS.

Alternatively, the path planning issues of UAVs are addressed to satisfy the constantly changing user demand.
In a three-layer edge computing architecture \cite{Ref93}, before the Lyapunov optimized online task offloading from UAVs to cloud servers, DRL-based path planning is first adopted to enable UAVs to collect sufficient data from distributed IoT devices.
When a UAV is flying over multiple mobile devices, a QoS-based offloading policy is developed by the double DQN algorithm \cite{Ref96}, with the goal of maximizing the system reward under the constraints of a minimal number of offloaded tasks and total energy consumption.
The joint optimization of user association, resource allocation, and UAV trajectory was investigated in \cite{9354996}, aiming at minimizing the energy consumption of all mobile devices.
Owing to the high complexity of conventional convex optimization algorithms, the DDPG algorithm is applied to make real-time trajectory control, while user association and resource allocation decisions are made by a matching algorithm.
In the multi-UAV-aided MEC system, UAV deployment is another optimization dimension, which has an important effect on the communication performance and load balance of the nonuniformly distributed ground IoT nodes.
Aiming at handling the unbalanced UAV workload issue during computational offloading, a differential evolution (DE)-based UAV deployment algorithm was first proposed, and then task scheduling was performed by a DRL algorithm \cite{Ref94}.
Furthermore, the objective in \cite{Ref95} was extended to maximize the fairness of the workload of each UAV and the number of service times for each mobile device while minimizing the overall energy consumption of all mobile devices, and the MADDPG algorithm was proposed to address the concern problems.
In large-scale UAV-enabled networks, \cite{9397778} considered a two-layered optimization problem, which is hierarchically decomposed into trajectory optimization and offloading decision optimization.
Based on the designed hierarchical RL, the DDPG and DQN algorithms are employed to address the two subproblems with low computational complexity and improved learning efficiency.

\begin{table*}[!t]
\centering
\caption{Comparisons of RL-Based Vehicular MEC Server and UAV-Based MEC Server}
\label{tab:tab8}
\begin{tabular}{|m{0.15in}|c|c|c|m{0.4in}|c|c|c|c|m{0.3in}|m{0.5in}|m{0.65in}|m{0.3in}|m{0.55in}|}
\hline
\multirow{2}{*}{\textbf{The.}} & \multirow{2}{*}{\textbf{Ref.}} & \multicolumn{3}{c|}{\textbf{Objective}} & \multicolumn{3}{c|}{\textbf{Constraints}} & \multirow{2}{*}{\textbf{Sol.}} & \multicolumn{5}{c|}{\textbf{Reinforcement Learning}} \\ \cline{3-8} \cline{10-14}
 &  &
 \textbf{Delay} & \textbf{Energy} & \multicolumn{1}{c|}{\textbf{Other}} & \textbf{MD} & \textbf{CH} & \textbf{MEC} &
 &
 \textbf{Model} & \textbf{Algorithm} & \multicolumn{1}{c|}{\textbf{MEC State}} & \textbf{Action} & \multicolumn{1}{c|}{\textbf{Reward}}  \\ \hline
\multicolumn{1}{|c|}{\multirow{3}{*}{\begin{sideways} Vehicle  \end{sideways}}} &   \cite{Ref91}    & \multicolumn{3}{m{1.4in}|}{Priced bits of transmission and computation}  &  \checkmark  & \checkmark  & \checkmark  & S1 &   MDP  &  {\it Q}-learning, DRL  &  Available server, computational resource  & \ding{172} \ding{182} \ding{183} &  Long-term utility   \\  \cline{2-14}
 & \cite{9421372} & \checkmark  & ----- & \multicolumn{1}{c|}{-----} & ----- & ----- & ----- & S1 & MDP & Multiagent DRL & Congestion level, traffic light & {\tiny\textcircled{\tiny{\textbf{12}}}} &  Reward of average vehicle flow control   \\  \cline{2-14}  
 & \cite{9336659} & \checkmark & ----- & Reputation value  & \checkmark & \checkmark & \checkmark & S1  & Multi-agent MDP  & Multiagent DRL & Reputation value, resource, velocity, distance   & \ding{172}  & System utility  \\ \hline
\multicolumn{1}{|c|}{\multirow{12}{*}{\begin{sideways} UAV  \end{sideways}}}   &   \cite{Ref93}   &  ----- & ----- & Collected data bits  &  \checkmark & \checkmark & \checkmark & S2  & MDP  &   DRL + Lyapunov optimization &  Position, data freshness  & \ding{179}  & Collected data bits  \\ \cline{2-14}
  & \cite{Ref96}   & ----- & \checkmark & Offloaded task number &  \checkmark & \checkmark & \checkmark  & S1 &  MDP   &   Double DQN   & Location, available energy &  \ding{179}  &  System reward  \\ \cline{2-14}
 & \cite{9354996} &  ----- & \checkmark & \multicolumn{1}{c|}{-----} & -----& -----& ----- & S2 & \multicolumn{1}{c|}{-----} & DDPG + matching algorithm & Location & \ding{179} & Negative energy consumption plus penalty \\ \cline{2-14}
  &   \cite{Ref95}   &  ----- & \checkmark & Fairness  &  \checkmark & \checkmark & \checkmark   & S1  & Multi-agent MDP   &  MADDPG  &  Coordinates, distance, service time, workload &  \ding{179}  &  Objective function minus flying penalty  \\ \cline{2-14}
& \cite{9397778} & \checkmark & ----- &  \multicolumn{1}{c|}{-----} & \checkmark & \checkmark & \checkmark & S2 & MDP &  DDPG + DQN & Location, energy consumption & \ding{172} \ding{175} & Reciprocal of delay  \\ \cline{2-14}
& \cite{Ref98}  &  \checkmark & \checkmark & Bandwidth  & \checkmark & \checkmark & \checkmark  & S1 &  MDP  & Double DQN  &  Energy queue &  \ding{172}  & Cumulative reward \\ \cline{2-14}
 &  \cite{Ref92}  &   \checkmark & \checkmark & \multicolumn{1}{c|}{-----}   &  -----&  -----&  -----  &  S2    &  Game, MDP  & Coalition formation + RL   & Backlog   &  \ding{172} \ding{175}   &   Reciprocal of cost   \\ \cline{2-14}
  &  \cite{Ref99}  & ----- &  \checkmark  & Forwarding and processing costs   & \checkmark & ----- & -----   &S2 &  MDP/ POMDP   &  RL/DRL + Bayesian RL/DL & Deadline, location, servers' action &  \ding{174}\ding{175} / \ding{172}  &   Payoff    \\ \cline{2-14}
  &   \cite{Ref94} &  \checkmark & -----  & load balance degree & \checkmark & \checkmark & \checkmark  & S2 & \multicolumn{1}{c|}{-----}  &  DQN+DE algorithm & Computational resource, execution time  & \ding{173}  &  Negative reciprocal of completion time  \\ \cline{2-14}
& \cite{Ref97}  &   \checkmark &  \checkmark  & \multicolumn{1}{c|}{-----} & \checkmark & \checkmark & \checkmark &  S1    &  MDP  &  {\it Q}-learning   & Delay, energy consumption  & \ding{172} &  Reciprocal of delay  \\ \cline{2-14}
& \cite{9366889}& \checkmark &  \checkmark  & \multicolumn{1}{c|}{-----} & \checkmark & \checkmark & \checkmark & S1 & MDP & DDPG &  Computational capacity, deadline, UAV controller & \ding{172} \ding{182} \ding{185} & Negative cost plus buffer overflow penalty \\ \cline{2-14}
 &  \cite{Ref134}   &  \multicolumn{3}{m{1.4in}|}{Priced bits of transmission and computations}   & \checkmark & \checkmark & \checkmark  & S1  & MDP &  DRL  & Size of computational demands & \ding{172} \ding{182}  &  System utility  \\ \hline
 {\begin{sideways} Hybrid \end{sideways}}  & \cite{9426913} & ----- & ----- & AoI & \checkmark & \checkmark & \checkmark  & S2 & MDP & FL+ multiagent DRL & Computational resource & \ding{172} \ding{179} \ding{182}  & Averaged AoI \\ \hline
\end{tabular}
\end{table*}

In addition, coexisting ground vehicles and UAVs will create more complicated network dynamics in MEC.
In \cite{9426913}, a joint optimization problem of trajectory planning, task scheduling, and resource allocation was investigated for system timeliness, i.e., the freshness of data and computational tasks.
The concept of AoI in \cite{Ref53} is used to evaluate the content freshness, which is expressed as the time of the current content elapsed since it is generated.
To minimize the AoI, policy-based multiagent DRL algorithms are employed to handle the high-dimensional action space.
Moreover, to improve communication efficiency and learning convergence among multiple agents, the FL mode is integrated into multiagent coordination.
Detailed comparisons of works on mobile MEC servers are listed in Table \ref{tab:tab8}.

\subsubsection{Server Cooperation}
Challenged by various task requirements and diverse radio access modes, it is unaffordable to support large requested tasks in a single MEC server.
Hence, there is a need to offload computation tasks from the current server to its adjacent MEC servers and cloud server for cooperation.
In the existing works, many RL algorithms have been developed for server cooperation \cite{Ref120,Ref121,Ref69,Ref123,Ref122,Ref70}.
To avoid overloading the MEC server, extra offloaded tasks are forwarded from edge computing to cloud computing under the control of an SDN-based module \cite{Ref120}.
Upon monitoring task dynamics and MEC servers' resources, the SDN-based controller enables a {\it Q}-learning-based policy of task offloading and resource allocation.
To accelerate the search speed of the {\it Q}-learning algorithm, based on multiagent RL, a cooperative {\it Q}-learning algorithm was proposed in \cite{Ref120}, where new agents can obtain efficient training and learning according to the shared information of experienced agents.
Without MEC server forwarding, i.e., the tasks can be directly offloaded to the cloud or the MEC, \cite{Ref123} minimized system energy consumption by task offloading and resource allocation based on cloud-edge-terminal orchestration.
Attributed to the mixed-integer problem, the DDPG algorithm is employed with the mapping from continuous-valued actions to refined discrete-valued actions by the constructed bipartite graph.
In \cite{Ref121}, a multilayer MEC was introduced for cloud-edge-terminal cooperation, where the remaining raw data can be forwarded from the bottom MEC server to its parent counterpart and then to the cloud.
To support a low-delay service with heterogeneous resources across multiple layers, the joint optimization of task offloading, radio spectrum access, pricing mechanism design, and network congestion control is proposed.
Due to the complex coupling between resource allocation and task offloading with time-varying task generation and unstable wireless channels, the {\it Q}-learning algorithm is adopted by the agent located in the cloud center for cloud-edge-terminal cooperation.
To satisfy the online QoS requirements in terms of delay and energy consumption, multiobjective optimization was formulated for server cooperation in \cite{Ref69} to minimize the weighted sum of computational time, transmission time, and energy consumption under bandwidth resource constraints.
By observing the bandwidth dynamics in the uplink and downlink, the optimal offloading location (i.e., the associated MEC server) is achieved by the SARSA-based DRL algorithm.
Simulation results show that the proposed algorithm outperforms the conventional {\it Q}-learning algorithm.

To simplify server cooperation and improve offloading efficiency, the system loss minimization problem was decoupled into two subproblems in \cite{Ref122}: resource allocation and offloading decision-making.
The former is first optimized by the quasiconvex bisection and polynomial analysis algorithms, and then the latter is solved by the RL algorithm.
Based on this, future resource utilization can be predicted, and the offloaded tasks can be properly divided into two parts for high-efficiency parallel edge-cloud computing.
Owing to the free mobility of mobile devices, the multitask offloading delay significantly increases, when mobile devices travel through hotspot areas with overloaded MEC servers.
To this end, parallel offloading to a macro BS and a small BS is presented in \cite{Ref70}, where each BS is equipped with one MEC server.
Upon partitioning the coverage of the macro BS into multiple small areas, different mobile devices in the same small area have the same offloading pattern (server selection and offloaded data size), and independent offloading patterns exist in different small areas.
Upon minimizing the execution time of all tasks, the optimal cooperative offloading pattern is achieved by the DDPG-based RL algorithm.
The studies on collaborative MEC servers are compared in Table \ref{tab:tab3}.

\subsubsection{Service Migration}
Service migration always occurs due to user mobility and workload dynamics in the MEC server.
To evaluate the service migration cost caused by user mobility in an unpredictable MEC environment, the authors of \cite{Ref67} propose a digital twin (DT)-aided MEC, where a digital representative similar to the real MEC environment is estimated by DT technology.
To evaluate the effect of the increased number of mobile devices and the MEC environment information (e.g., mobile device trajectory and MEC server workload),
the average offloading delay during the entire trip of the mobile device is formulated as the objective function, subject to the service migration cost, maximum task delay, and available MEC servers.
Then, similar to \cite{9348135,9262878}, upon leveraging the Lyapunov optimization, the original optimization problem is simplified as a multiobjective dynamic optimization problem.
Based on the actor-critic  DRL algorithm, optimal MEC server selection is achieved with low offloading delay and failure rate of the computation tasks.
In \cite{Ref77}, accounting for the free mobility and limited observations among distributed mobile devices, the average task completion delay may be increased under the constraint of the migration energy budget.
Based on the multiagent DRL mechanism, a centralized critic network in the MEC server collects the reward of each mobile device and shares the global reward to distributed actor networks in mobile devices.
It is demonstrated that the proposed algorithm has a much lower average completion time than the algorithms with a single-agent actor-critic  network or without service migration.
By considering different speeds and varying service requirements of mobile users, speed-aware task offloading and resource allocation were further developed by the advantage actor-critic-based RL algorithm in \cite{9416470}.
According to speed and tolerable delay, priorities are assigned to mobile devices with different offloading requests to conduct efficient resource allocation and task offloading.
For secure computational offloading, the blockchain-empowered task migration optimization problem was investigated in \cite{9444665}.
According to the Lyapunov optimization, the original problem is simplified as online QoS maximization without the long-term handover cost, followed by the actor-critic DRL-based solution.
The QoS is measured by the difference between the required delay and the actual delay.
The detailed comparisons are shown in Table \ref{tab:tab3}.

\begin{table*}[thpb]
\centering
\caption{Comparisons of RL-Based MEC with Server Cooperation and Service Migration}
\label{tab:tab3}
\begin{tabular}{|m{0.15in}|c|c|c|c|c|c|c|c|m{0.3in}|m{0.6in}|m{0.6in}|m{0.35in}|m{0.6in}|}
\hline
\multirow{2}{*}{\textbf{The.}} & \multirow{2}{*}{\textbf{Ref.}} & \multicolumn{3}{c|}{\textbf{Objective}} & \multicolumn{3}{c|}{\textbf{Constraints}} & \multirow{2}{*}{\textbf{Sol.}} & \multicolumn{5}{c|}{\textbf{Reinforcement Learning}} \\ \cline{3-8} \cline{10-14}
 &  &
 \textbf{Delay} & \textbf{Energy} & \textbf{Other} & \textbf{MD} & \textbf{CH} & \textbf{MEC} &
 &
 \textbf{Model} & \multicolumn{1}{c|}{\textbf{Algorithm}} & \multicolumn{1}{c|}{\textbf{MEC State}} & \multicolumn{1}{c|}{\textbf{Action}} & \multicolumn{1}{c|}{\textbf{Reward}}  \\ \hline
\multicolumn{1}{|c|}{\multirow{6}{*}{\begin{sideways} Server Cooperation  \end{sideways}}}
    &  \cite{Ref123}   & ----- & \checkmark & ----- & \checkmark & \checkmark & \checkmark & S1  & MDP & DDPG & Computational resource, deadline  & \ding{172} \ding{182}  & Negative system energy consumption  \\ \cline{2-14}
    &   \cite{Ref122}  &\checkmark & \checkmark & -----   &   \checkmark & \checkmark & \checkmark  & S2  & MDP &  Quasiconvex bisection + polynomial analysis + {\it Q}-learning  & Computational resource & \ding{182} &  Negative loss ratio sum of delay and energy  \\ \cline{2-14}
     & \cite{Ref70}  &  \checkmark & ----- & ----- &  \checkmark & \checkmark & \checkmark &   S1   &   MDP   &   DDPG   &   Server queue     &    \ding{172} \ding{175}   &  Negative total delay  \\ \cline{2-14}
    &  \cite{Ref69}   &  \checkmark & \checkmark & -----  & ----- &  \checkmark & ----- &   S1   &  MDP  &  SARSA    &  CPU frequency, memory & \ding{172}   & Resource allocation reward  \\ \cline{2-14}   
     &  \cite{Ref120} & \checkmark & -----  & ----- & \checkmark & \checkmark & \checkmark  & S1  & \multicolumn{1}{c|}{-----} &  {\it Q}-learning, multiagent RL  & System delay, CPU frequency  & \ding{182} & Reciprocal of minimal delay   \\ \cline{2-14}
   &   \cite{Ref121}   &  \checkmark & -----  & -----   &  -----&  -----&  -----    & S1  & \multicolumn{1}{c|}{-----} &  {\it Q}-learning  & \multicolumn{1}{c|}{-----}  & \ding{172} \ding{182} \ding{183} \ding{184} \ding{185} & Function of delay   \\ \hline
\multicolumn{1}{|c|}{\multirow{4}{*}{\begin{sideways} Service Migration \end{sideways}}}    &    \cite{Ref77}  &  \checkmark & -----  & -----   &   \checkmark & \checkmark & \checkmark   & S1  &  MDP &  Multiagent DRL  &  CPU frequency, previous and current serving node &  \ding{175} & Average completion time difference \\ \cline{2-14}
      &     \cite{Ref67}     &  \checkmark & -----  & -----  & \checkmark & \checkmark & \checkmark  & S1  & MDP   & Lyapunov optimization, actor-critic  DRL  &   CPU frequency, migration cost    &  \ding{175}  &   Offloading delay, migration cost     \\ \cline{2-14}
     & \cite{9416470} & \checkmark & -----  & ----- & \checkmark & \checkmark & \checkmark & S1 & MDP & Advantage actor-critic DRL & Offloaded server & \ding{172} & Negative delay  \\ \cline{2-14}
     & \cite{9444665}& \checkmark & -----  & ----- & \checkmark & \checkmark & \checkmark & S1 & MDP & Actor-critic  DRL & Server load, migration cost & \ding{175} & Delay difference minus migration cost \\ \hline
\end{tabular}
\end{table*}

\subsubsection{Learned Lessons}
As shown in Tables \ref{tab:tab4}, \ref{tab:tab8}, and \ref{tab:tab3}, the CPU, storage, and location states at the MEC side are always observed to assist agents in performing computation- and caching-aware RL.
Since the well-trained DRL agent requires numerous interactions with the time-varying MEC environments to obtain a large amount of experienced data with high diversity, trial-and-error will cause a considerable exploration cost, which is unaffordable for a single MEC agent (single environment).
In this regard, multiple MEC agents (i.e., multiple environments) are necessary to make intelligent decisions in MEC-based networks.
For instance in \cite{Ref76} with multiple environments, the transition states involve task information, mobile device, channel information, and MEC server status.
In detail, data size, required CPU cycles, and tolerable task delay are in the task information, CPU capacity, task queuing delay, location are the statuses of mobile devices, channel gains and noise are included in the channel information, and CPU capacity and task queuing delay are considered in the MEC server.
Upon capturing diverse states, the optimal offloading decision and channel selection are achieved by a decentralized DRL-based algorithm.
However, in multiple environments, the high communication efficiency, robustness against non-i.i.d. data distribution, and scalability for collective training are required for the distributed MEC agents.
By collecting experiences of distributed agents, a master agent can share a global experience with all distributed agents to conduct efficient training.
Furthermore, based on adaptive $n$-step learning, efficient training with faster convergence and smaller variance is obtained compared to the conventional distributed DRL algorithm.
However, in different regions covered by a large-scale network, multiple master agents may be required to collect dynamic environmental information and share the global characteristic model.
In this regard, how to deploy and manage these master agents (distributed, centralized, or multilayer hybrid) in MEC-based networks for communication-efficient RL is still an open problem.

\section{State-of-the-Art Research: A Perspective of Diverse Applications}

In this section, we investigate RL-empowered MEC to support diverse applications in uncertain network environments.
Specifically, the concern applications include multimedia applications, industrial applications, vehicular applications, IoT applications, and trust-based applications.

\subsection{Multimedia Applications}
Multimedia applications include immersive VR applications, online gaming, video applications, and so on.
As a common point, a high QoE of mobile users is expected for multimedia applications. In this regard, computation offloading and content caching in MEC are perceived as promising ways.
In \cite{Ref82,Ref83}, the offloading policy through terahertz (THz) communication was investigated to support immersive VR applications.
Based on the popularity and correlation of the rendered contents, an adaptive VR framework including local, remote, and cooperative rendering was proposed in \cite{Ref82} to provide high-quality and real-time rendering.
In the proposed framework, DRL-enabled offline training and game theory-based online running are developed to maximize the QoE of mobile users with Nash equilibrium.
In \cite{Ref83}, considering the time-varying characteristics of the THz channel and the mobility of the mobile device, the viewport renderings at the head-mounted display and MEC server were jointly optimized to minimize the energy consumption of the mobile device.
Via minimizing the long-term energy consumption, the A3C-based DRL algorithm is designed to achieve the optimal policy of viewport rendering offloading and downlink transmit power control.
To enhance the QoE of interactive VR applications, decoupled learning was employed in \cite{9411714} for the field of view (FoV) prediction and MEC-enabled FoV rendering.
The former is captured by the RNN at the central controller.
Then, based on the location correlation and predicted FoV, the latter is optimized by the centralized and distributed DRL to make the optimal user association and rendering model migration.
To overcome the challenges of mobile devices running large-resource-consuming games, i.e., low-delay requirement, high bandwidth consumption, and sensitive users' QoE,
\cite{Ref84} leveraged the computation and caching resources in the MEC server for game-rendering offloading.
Moreover, a DRL-based adaptive bit rate method is developed to guarantee users' QoE in dynamic environments.

MEC systems are also applied to diverse video applications, such as short video \cite{Ref144}, 3D video \cite{Ref146}, and video surveillance \cite{Ref147}.
In these applications, many feasible technologies can be applied in MEC systems, such as radio bearer control, video quality selection, adaptive bitrate streaming, video transcoding, video playback-based caching, face recognition methods, and object detection.
In \cite{Ref144}, the joint optimization of video quality selection and radio bearer control was proposed to obtain the maximal video service profit while maintaining the minimal radio bearer's cost and timeout penalty.
Due to the large state-action space size caused by the dynamic setup, reconfiguration, and release in the radio bearer controller, a policy-based DRL algorithm is adopted to achieve the optimal solution for short video transmission.
In \cite{Ref52}, a new framework of joint video transcoding and quality adaptation was presented for adaptive bitrate streaming in time-varying wireless channels.
The optimization objective function is defined as the time-averaged network reward, which is the weighted sum of the video quality, video quality variation, video rebuffering time, playback penalty, transcoding penalty, and CPU consumption.
The first three terms form the QoE and the remaining terms are related to the transcoding cost.
Then, the double DQN algorithm is adopted to adaptively assign the computational resources and adjust the video quality.
In \cite{Ref146}, an MDP-based rate adaptation scheme was proposed for the block transmission speed assignment of 3D video streaming in a time-varying MEC environment.
Based on the predicted bandwidth and viewport via the LSTM network and the observed historical block information, the actor-critic-based DRL algorithm is applied for differential transmission in video playback.
It is demonstrated that the QoE is improved through adaptive rate allocation for video tiles.
For video surveillance, video offloading technology is always adopted to enrich the user's experience \cite{Ref147,9406368}.
In detail, by jointly optimizing video offloading, bandwidth allocation, and image compression, an RL-based image recognition algorithm was proposed in \cite{Ref147} for video surveillance.
In \cite{9406368}, to minimize the processing time and maintain the recognition accuracy, a two-layer machine learning framework was proposed by combining DQN with supervised learning.
The optimal strategy of offloading and bandwidth allocation is based on DQN, while neural network-based supervised learning is used to select the compression rate.
The detailed comparisons are shown in Table \ref{tab:tab5}.

\subsection{Industrial Applications}

In industrial applications, based on digital technologies (e.g., DT \cite{Ref86} and blockchain \cite{Ref118}), multiple machine-type communication devices are involved in smart manufacturing, environment monitoring, data analytics, automatic control, intelligent grids, etc.
Attributed to massive data transmission and processing, low delay and high reliability are expected in industrial applications.
In this regard, MEC-empowered multichannel access and task offloading have become pivotal points for a high QoS.
Due to the dynamics and continuity of task generation in the industrial network, continuous action and state require high-dimensional spaces.
Thus, DDPG-based resource allocation was proposed in \cite{Ref117} to minimize the long-term average task delay.
In \cite{Ref85}, the joint control of wireless channel access and task offloading was designed in an industrial network.
For highly efficient spectrum utilization, D2D communication is introduced, where each device corresponds to a single agent.
To make the optimal control of channel access and computational delay in the multiagent environment, based on the POMDP for the collaborative agents, the MADDPG-based DRL algorithm is proposed in \cite{Ref85}.
Furthermore, stochastic tasks, heterogeneous resources, and resource-constrained industrial mobile devices were investigated in \cite{Ref86}.
To minimize the long-term energy efficiency, similar to \cite{Ref67}, Lyapunov optimization is exploited to transform the long-term stochastic computation offloading problem into a deterministic per-time block problem.
With DT-enabled real-time monitoring for mobile devices and BSs, the optimal policy of computational offloading and resource allocation is made by an asynchronous actor-critic (AAC)-based DRL algorithm.
In \cite{Ref118}, a blockchain-enabled industrial IoT network was investigated for data security and privacy.
In this regard, to solve the problems of unbearable energy consumption, poor efficiency of the consensus mechanism, and serious computational overhead, the MEC-based optimization framework is introduced.
By dynamically selecting the master controller, offloading decision, block size, and computing server, the DRL algorithm is applied to minimize the system cost (weighted sum of energy consumption and computation overhead).
The research on industrial applications is summarized in Table \ref{tab:tab5}.

\begin{table*}[!t]
\centering
\caption{Comparisons of RL-Based MEC in Multimedia and Industrial Applications}
\label{tab:tab5}
\begin{tabular}{|m{0.15in}|c|m{0.65in}|m{0.5in}|m{0.1in}|m{0.3in}|m{0.5in}|m{0.6in}|m{0.45in}|m{0.5in}|m{0.3in}|m{0.6in}|}
\hline
\multirow{3}{*}{\textbf{The.}} & \multirow{3}{*}{\textbf{Ref.}} & \multicolumn{1}{c|}{\multirow{3}{*}{\textbf{Objective}}} & \multirow{3}{*}{\textbf{Constraints}} & \multirow{3}{*}{\textbf{Sol.}} & \multicolumn{7}{c|}{\textbf{Reinforcement Learning}} \\  \cline{6-12}
 &  &    & & & \multirow{2}{*}{\textbf{Model}} & \multirow{2}{*}{\textbf{Algorithm}} & \multicolumn{3}{c|}{\textbf{States}} & \multirow{2}{*}{\textbf{Action}} & \multicolumn{1}{c|}{\multirow{2}{*}{\textbf{Reward}}}  \\ \cline{8-10}
  &  &   &  &   & & & \multicolumn{1}{c|}{\textbf{Mobility}} & \multicolumn{1}{c|}{\textbf{Channel}} & \multicolumn{1}{c|}{\textbf{MEC}} &  &   \\ \hline
\multicolumn{1}{|c|}{ \multirow{9}{*}{\begin{sideways}Multimedia applications\end{sideways}}}
   &    \cite{Ref82}    &  Delay, energy   & MD, CH, MEC  &  S2   & \multicolumn{1}{c|}{-----} & Double dueling DQN+game  & User request, user association & Distance & Caching  & \ding{172} \ding{175} \ding{184}  &  Long-term averaged utility   \\ \cline{2-12}
 &    \cite{Ref83}    &  Energy   &  MD, CH  &   S1   &   MDP &  A3C  &  Task queue & Distance, path loss, subwindow number & Task queue  &  \ding{172} \ding{185}   &  Reciprocal of energy consumption \\ \cline{2-12}
 & \cite{9411714} & Peak SNR & MD, CH, MEC & S2 & POMDP & RNN+DRL & FoV & Distance & Computa-tional resource & \ding{175} \ding{181} & Peak SNR  \\ \cline{2-12} 
  & \cite{Ref84} & Video bitrate, smoothness, rebuffering  &  \multicolumn{1}{c|}{-----}  &  S1  & \multicolumn{1}{c|}{-----} &  A3C  & NACK count, jitter buffer level & Packet loss rate, received bitrate & Bitrate, buffer level &  \ding{180}  & QoE  \\ \cline{2-12}
 & \cite{Ref144}   &  Quality profit, delay penalty, radio bearer cost  &  MD, CH, MEC & S1  & MDP & DRL & Video request & Channel quantity, transmission rate & Waiting time, video quality level  & \ding{177} \ding{183} & Video reward    \\ \cline{2-12}
& \cite{Ref146}  &  Buffering time, quality fluctuation, average quality   &  \multicolumn{1}{c|}{-----}  & S2  & MDP & LSTM + double DQN   & Viewport & Bandwidth & Buffer occupancy, tile size  & \ding{180} & QoE  \\ \cline{2-12}
 &\cite{Ref147}   &  Delay   &  \multicolumn{1}{c|}{-----}  & S1  &  \multicolumn{1}{c|}{-----} & {\it Q}-learning   & Confidence level & CSI & \multicolumn{1}{c|}{-----}  & \ding{172} \ding{180} &  Reciprocal of time function  \\ \cline{2-12}
&\cite{Ref52}   &  Video quality, rebuffering time, penalties  &  MD, CH, MEC & S1  & MDP & Double DQN  & Playback buffer & CSI & Transcoder buffer, video quality & \ding{180} &  QoE and transcoding cost  \\ \cline{2-12}
  & \cite{9406368} & \multicolumn{1}{c|}{-----} & \multicolumn{1}{c|}{-----} & S2 & MDP & DQN + supervised learning & Image size, confidence level, waiting time & CSI, channel occupation & \multicolumn{1}{c|}{-----} & \ding{172} \ding{180} \ding{183}  & Negative exponent of delay \\ \hline
\multicolumn{1}{|c|}{\multirow{4}{*}{\begin{sideways} Industrial applications \end{sideways}}}
      &  \cite{Ref86}  &   Energy   &  MD, CH, MEC  & S2 &  MDP &  Lyapunov optimization+AAC  & Computational resource, transmit power, task queue & Transmis-sion rate, bandwidth & Computa-tional resource, task queue  & \ding{182} \ding{183} \ding{184} \ding{185} & Function of energy consumption  \\ \cline{2-12}
       &  \cite{Ref118}   & Energy, computation overhead &   \multicolumn{1}{c|}{-----}   &  S1 & MDP &  DQN  & \multicolumn{1}{c|}{-----} & \multicolumn{1}{c|}{-----} & Energy level, computational resource and overhead  &  \ding{172} \ding{175} \ding{176} &  Consensus reward minuses cost and punishment  \\ \cline{2-12}
      &  \cite{Ref117}   &  Delay & MD, CH, MEC  & S1  & MDP & DDPG   & CPU cycle, data size, battery level &  CSI & \multicolumn{1}{c|}{-----} &  \ding{182} \ding{185} &  Negative of average delay    \\ \cline{2-12}
      &   \cite{Ref85}    & \multicolumn{1}{c|}{-----}  &   \multicolumn{1}{c|}{-----}     &  S1   &   POMDP   &   MADDPG   & Data size, CPU cycle, task priority & CSI & \multicolumn{1}{c|}{-----} &  \ding{172} \ding{183}   & Transmission reward and delay difference  \\ \hline
\end{tabular}
\end{table*}

\subsection{Vehicular Applications}
Due to high-speed vehicle mobility, the wireless channel states change rapidly, which creates high system performance requirements, such as ultralow delay and low handover cost.
In this regard, computational offloading, server cooperation, service migration, path planning, and content caching are used for low-latency service provisioning \cite{Ref79,Ref80,Ref81,Ref128,Ref129,Ref119,Ref138,9366426,Ref158}.
To handle where and when to take offloading actions in MEC-based vehicular networks with diverse task characteristics, time-varying wireless environments, and frequent handovers, the complicated stochastic optimization problem is formulated as the MDP in \cite{Ref138}.
To improve the training efficiency and accelerate the convergence speed of the DRL algorithm with enormous states, the proximal policy optimization algorithm in the policy gradient method is first proposed.
Then, the offloading policy and value function are evaluated by a parameter-shared DNN to extract representative features.
Finally, efficient exploration attempts are made during the training by the deliberate representation adjustment of states and rewards.
In \cite{Ref79}, an adaptive task offloading policy is proposed based on the DDPG algorithm to minimize the system cost in a rapidly varying environment.
Numerical results show that the proposed algorithm can achieve better efficiency than the dueling DQN and greed policy.

To provide low-latency and reliable offloading services, \cite{Ref80} focused on a cooperative offloading problem by balancing the communication capacity and computational overhead.
To handle the optimization complexity issue in the time-varying network topology, the high-dimensional optimization problem is divided into two subproblems.
The first problem focuses on task partitioning and scheduling among MEC servers without cooperative transmission based on a heuristic algorithm to improve service reliability.
The second problem is addressed by the DDPG-based DRL algorithm to select the MEC server and obtain the optimal offloading policy.
Considering server cooperation, \cite{Ref119} proposed a regional intelligent computational resource allocation for tasks with different delay tolerances using the DQN algorithm.
By separating the vehicular network into multiple regions, several MEC servers in the same region are managed by a centralized MEC region controller.
Thus, the overloaded tasks for a single MEC server can be processed by the cooperation of several MEC servers.
Ref. \cite{Ref128} considered a cloud-edge-terminal vehicular network with D2D communication.
To reduce the task delay, a multiplatform intelligent algorithm of offloading and resource allocation is proposed in dynamic vehicular networks.
First, the $K$-nearest neighbor (KNN)-based classification algorithm is utilized to determine the platform (local computing, vehicular edge computing, or cloud computing) for low-latency offloading.
Then, the cooperative resource allocation during task offloading is optimized by the {\it Q}-learning algorithm.
In \cite{Ref81}, using the social relationship of vehicle owners, the authors proposed a joint optimization of computing and caching for content offloading and dispatch.
Due to the diverse social characteristics and cross-area coupling in the vehicular network, a DDPG-based content distribution strategy is learned when the vehicle travels across multiple areas.
In \cite{Ref129}, the joint optimization of migration decision, offloading policy, and resource allocation was formulated, considering the time-varying wireless CSI, stochastic task arrival, bandwidth allocation, and hosting caching.
The double DQN algorithm is adopted to minimize the total cost of delay, energy consumption, bandwidth allocation, and timeout penalty.

\begin{table*}[hptb]
\centering
\caption{Comparisons of RL-Based MEC in Vehicular Applications}
\label{tab:tab6}
\begin{tabular}{|c|m{0.7in}|m{0.5in}|c|m{0.25in}|m{0.5in}|m{0.65in}|m{0.5in}|m{0.6in}|m{0.3in}|m{0.6in}|}
\hline
 \multirow{3}{*}{\textbf{Ref.}} & \multicolumn{1}{c|}{\multirow{3}{*}{\textbf{Objective}}} & \multirow{3}{*}{\textbf{Constraints}} & \multirow{3}{*}{\textbf{Sol.}} & \multicolumn{7}{c|}{\textbf{Reinforcement Learning}} \\  \cline{5-11}
   &    & & & \multirow{2}{*}{\textbf{Model}} & \multirow{2}{*}{\textbf{Algorithm}} & \multicolumn{3}{c|}{\textbf{States}} & \multirow{2}{*}{\textbf{Action}} & \multicolumn{1}{c|}{\multirow{2}{*}{\textbf{Reward}}}  \\ \cline{7-9}
    &   &  &   & & & \multicolumn{1}{c|}{\textbf{Mobility}} & \multicolumn{1}{c|}{\textbf{Channel}} & \multicolumn{1}{c|}{\textbf{MEC}} &  &   \\ \hline
    \cite{Ref80}     &   Delay, service failure penalty  &  MD, CH, MEC  &   S2   &  MDP  &  Heuristic algorithm + DDPG  &  Data amount, vehicle speed & \multicolumn{1}{c|}{-----} & Completion delay  &  \ding{175}  &  Discounted cost  \\ \hline
    \cite{Ref79}   &   Delay, energy, bandwidth  & MD, CH, MEC &   S1    & \multicolumn{1}{c|}{-----} & DDPG   &  Task size & CSI, SINR & Buffer queue length  &   \ding{183} \ding{185}  &   Negative average cost  \\ \hline
   \cite{Ref81}  &  Number of vehicles with dispatched content & MD, CH, MEC  &  S1    &  \multicolumn{1}{c|}{-----}   &  DDPG   &  \multicolumn{3}{m{2.1in}|}{Number of vehicles with dispatched content} & \ding{172} \ding{173} \ding{184} & Average objective function         \\ \hline
  \cite{Ref128}   &  Delay   &  MD, CH, MEC &  S2 & MDP & KNN + {\it Q}-learning   & Local delay & Transmission delay & Computing delay, computational resource &   \ding{172} \ding{182} & Negative total delay  \\ \hline
  \cite{Ref129}   &  Delay, energy, bandwidth   & MD, CH, MEC & S1 & \multicolumn{1}{c|}{-----} & Double DQN & Task size, CPU cycle, deadline & CSI & Migration decision  & \ding{172} \ding{181} \ding{183} &  Negative total cost  \\ \hline
   \cite{Ref119}   &   Delay   & \multicolumn{1}{c|}{-----}  & S1  & MDP &  DQN  &  Task type & \multicolumn{1}{c|}{-----} & Workload, computational resource & \ding{175} \ding{182} & Reciprocal of delay, negative delay \\ \hline
   \cite{Ref138} &  Delay, energy & \multicolumn{1}{c|}{-----} & S1  & MDP &  DRL  &   Task queue, computational resource, transmission queue  & Location, distance & Connected MEC server  & \ding{172} \ding{173} & Negative cost \\ \hline
    \cite{9366426} & Traveled distance & MD, CH, MEC & S1 & MAB  & DQN & \multicolumn{3}{m{2.1in}|}{Carryover component, traveled distance unit} & \ding{173} & Negative traveled distance  \\ \hline
    \cite{Ref158}   &  Fuel consumption   &  \multicolumn{1}{c|}{-----}  &  S1 & MDP &  DRL  & \multicolumn{1}{c|}{-----} & \multicolumn{1}{c|}{-----} & \multicolumn{1}{c|}{-----}  & \ding{179} &  Benefits of traveling time and distance  \\  \hline
\end{tabular}
\end{table*}

To support real-time computing services at a safe autonomous driving distance, in \cite{9366426}, distance-aware offloading and computational resource scheduling were investigated.
To guarantee that the vehicle can receive the computational results from the MEC server in time within a safe distance,
the objective is to minimize the traveled distance in an offloading duration, which is defined as the traveled distance from the end of receiving the previous offloading result to the beginning of receiving the current offloading result.
According to the mobility information of each vehicle, the restless MAB modeled stochastic scheduling problem for adaptive computational resource allocation in the MEC server is efficiently solved by the DQN algorithm.
Simulation results show that the computational results can be promptly delivered back to the vehicles by the proposed scheme in both synchronous and asynchronous offloading scenarios.
In \cite{Ref158}, with the aid of platooning technology, the path planning issue of autonomous vehicles was investigated to minimize the transport cost by reducing energy consumption and enhancing traffic efficiency.
With the task deadline and time-varying fuel consumption in the vehicular platoon, to maximize the saved fuel, the optimal path is planned by the DRL algorithm with known beginning and ending points.
The specific comparisons of vehicular applications are shown in Table \ref{tab:tab6}.

\subsection{General IoT Applications}

In addition to vehicular and industrial networks, IoT networks also include many specific scenarios, such as smart cities \cite{Ref114,Ref116,Ref157,8657791}, satellite-aided networks \cite{Ref131,9184934}, and narrowband (NB)-IoT \cite{Ref135,Ref136}.
For computation-intensive device access and delay-sensitive service requirements, MEC-based SDN, NFV, routing, and narrowband cellular transmission technologies provide centralized control, heterogeneous resource management, and effective communication.
In \cite{Ref114}, cloud-edge-terminal collaboration was considered in MEC-based smart city IoT networks.
In particular, a joint communication and computational resource allocation problem is formulated based on the Lyapunov drift plus penalty-based optimization theory to minimize the cost of total system delay subject to limited energy.
To solve the formulated problem effectively, it is first transformed into two subproblems: task offloading and task migration.
Then, the {\it Q}-learning-based task offloading algorithm crossing multiple network segments is designed.
Based on the Lagrange theorem, the task migration problem is solved for workload balance and resource utilization improvement among multiple MEC servers.
Based on \cite{8288203}, an integrated framework of networking, caching, and computing was proposed in \cite{Ref116} for smart cities.
To abstract, allocate, and optimize resources, the joint resource orchestration strategy is learned by the double dueling DQN algorithm with the aid of the SDN controller and NFV.
Owing to the dynamic crowd in various sectors of smart cities, the double DQN-based smart routing algorithm was devised to reduce network congestion and balance the network workload in \cite{Ref157}.
Furthermore, considering the high cost of infrastructure deployment and maintenance and the severe service pressure on MEC servers in cities, an SDN-enabled MEC architecture was proposed \cite{8657791}.
Then, the DQN-based RL algorithm is employed to minimize the service delay and balance the resource allocation.
The delay includes edge network routing delay and data processing delay, and the balance goal is to minimize the network load variance in each link and computation load variance at each MEC server.

The MEC-enhanced IoT network with multiple satellites and multiple satellite gateways was investigated in \cite{Ref131}, where the joint optimization of user association, offloading decision, and resource allocation was concerned.
In such a scenario, each satellite is responsible for processing data or forwarding data to gateways.
To minimize the weighted sum of system delay and energy consumption, similar to \cite{Ref122,Ref126}, two decomposed subproblems for resource allocation and the joint optimization of user association and offloading are optimized in turn.
The first subproblem is optimized by the Lagrange multiplier method, and the second subproblem is solved by the DQN algorithm.
Similarly, in \cite{9184934}, the joint optimization of task offloading and resource allocation in the satellite-based vehicle-to-vehicle (V2V) network was presented.
The optimization problem of delay minimization is also decomposed into a resource allocation subproblem and an offloading subproblem.
The former is solved by the Lagrange multiplier algorithm, and the latter is optimized by the model-free DRL algorithm.

To achieve efficient multidevice scheduling, in NB-IoT edge systems \cite{Ref135,Ref136}, a joint control policy of computational offloading and multiuser scheduling is formulated as the continuous-time MDP model to minimize the long-term average weighted sum of delay and power consumption with the aid of DRL algorithms.
In \cite{Ref135}, to solve the curse-of-dimensionality problem and reduce the signal overhead, an improved neural network architecture was designed for the value function approximation, including local feature summation for the global value function approximation, a convolutional layer for local state compression, and a multiplication layer for local value function updating.
In \cite{Ref136}, the linearly approximated value function and temporal-difference learning with postdecision state and semigradient descent were applied under uniformization.

Furthermore, a double auction game model was established to manage the resource by introducing an honest broker between MEC servers and IoT devices in \cite{Ref115}.
The objective is to maximize the utility of both MEC servers and IoT devices.
The utility of MEC servers is defined as the difference between the selling price of computational resources and the cost of construction and maintenance, while the utility of IoT devices is defined as the profit of a successful transaction.
To achieve the Nash equilibrium for bidding and asking policies among multiple participants (MEC servers and IoT devices), an experience-weighted attraction algorithm is proposed by combining RL with belief learning.
When a large quantity of sensor data is uploaded from IoT devices to the MEC server, to efficiently process the queued data, the task scheduling and computation resource allocation were jointly optimized in \cite{Ref112}.
To minimize the long-term weighted sum of the average task completion time and average number of requested computational resources, DQN-based resource allocation is proposed.
Three improvements were designed in \cite{Ref112} to improve the learning efficiency and training convergence of the DQN algorithm.
First, small mutual influence-based experiences are stored.
Second, through decoupling the decision epoch from the real timestep, the large-scale action space is divided into two small-scale action spaces, corresponding to task scheduling and resource allocation.
Finally, by introducing a filter layer at the end of the artificial neural network, the state-action values with invalid actions are removed.
In \cite{Ref113}, to achieve the efficient collaboration of task scheduling and resource allocation in the IoT network, a multitask DRL algorithm was proposed using self-play training to handle the issues of MEC server selection and wireless resource allocation.
Resource allocation is predicted by training a DNN in a self-supervised learning manner.
Based on this, by simulating multiple candidate trajectories of future states and actions, the MC tree search algorithm can achieve the optimal actions of offloading decision and resource allocation in MEC.
The detailed comparisons are listed in Table \ref{tab:tab7}.

\begin{table*}[thpb]
\centering
\caption{Comparisons of RL-Based MEC in General IoT Applications}
\label{tab:tab7}
\begin{tabular}{|m{0.2in}|m{0.7in}|m{0.5in}|c|m{0.25in}|m{0.5in}|m{0.6in}|m{0.5in}|m{0.65in}|m{0.35in}|m{0.6in}|}
\hline
 \multirow{3}{*}{\textbf{Ref.}} & \multicolumn{1}{c|}{\multirow{3}{*}{\textbf{Objective}}} & \multirow{3}{*}{\textbf{Constraints}} & \multirow{3}{*}{\textbf{Sol.}} & \multicolumn{7}{c|}{\textbf{Reinforcement Learning}} \\  \cline{5-11}
   &    & & & \multirow{2}{*}{\textbf{Model}} & \multirow{2}{*}{\textbf{Algorithm}} & \multicolumn{3}{c|}{\textbf{States}} & \multirow{2}{*}{\textbf{Action}} & \multicolumn{1}{c|}{\multirow{2}{*}{\textbf{Reward}}}  \\ \cline{7-9}
    &   &  &   & & & \multicolumn{1}{c|}{\textbf{Mobility}} & \multicolumn{1}{c|}{\textbf{Channel}} & \multicolumn{1}{c|}{\textbf{MEC}} &  &   \\ \hline
      \cite{Ref116}   &  SNR, computation capability, caching state   &   \multicolumn{1}{c|}{-----}   & S1  & MDP &  Double dueling DQN  & \multicolumn{1}{c|}{-----} & Available BS  & Available server, caching queue & \ding{172} \ding{175} \ding{184} &  Operator revenue  \\ \hline
      \cite{Ref131}   &  Delay, energy  &  CH, MEC   & S2  & MDP &  Lagrange Multiplier + DQN  & Arrival task, location & Communica-tion resource & Computational resource, location &  \ding{173} \ding{182} \ding{183} &  Negative objective function  \\ \hline
      \cite{9184934} &  Delay & MD, CH, MEC & S2 & MDP & Lagrange Multiplier + DQN & Unserved task, task size & Communica-tion resource & Computational resource, queuing delay  & \ding{172} & Decreasing function of delay \\ \hline
      \cite{Ref114}   &  Delay & MD, CH, MEC & S2  & MDP &  {\it Q}-learning + bisection search  & \multicolumn{3}{m{1.5in}|}{Offloading decision}  & \ding{172} &  Lyapunov function of delay and energy consumption \\ \hline
     \cite{Ref157}, \cite{8657791} & Delay & \multicolumn{1}{c|}{-----} & S1 & \multicolumn{1}{c|}{-----} & DQN & Request location & \multicolumn{1}{c|}{-----} & MEC location &  \ding{175}  &  Sum of delay and computation load   \\ \hline
     \cite{Ref135,Ref136} & Delay, energy & \multicolumn{1}{c|}{-----} & S1  & MDP & DRL & Task queue, behavior, and scheduling  & \multicolumn{1}{c|}{-----} & \multicolumn{1}{c|}{-----} & \ding{172} \ding{173} & Objective function \\ \hline
      \cite{Ref115}   &  Delay, computational resource   &  MD, MEC & S2  & Game & RL+belief learning & \multicolumn{3}{m{1.9in}|}{Experience number, attraction value}  & \ding{182} &  Reward of MD and MEC  \\ \hline
        \cite{Ref112}   & Delay, computational resource &  \multicolumn{1}{c|}{-----}  & S1  & MDP &  DQN  & Task size & Transmis-sion delay & Computational resource, task queue & \ding{173} \ding{182}  & Negative delay or computing units  \\ \hline
      \cite{Ref113} &  Delay/Energy & MD, CH, MEC & S1   & MDP &  DRL  & Task request & Bandwidth & Computational resource & \ding{172} \ding{182} \ding{183} & Average service delay \\ \hline
\end{tabular}
\end{table*}

\subsection{Trust-Based Applications}

In mobile networks, such as social networks \cite{Ref57,Ref58,8647548} and blockchain-based networks \cite{Ref87,Ref88,Ref89,Ref90,Ref132,Ref133,9385791,Ref160}, MEC is beneficial to trust-based communication among mobile devices due to its avoidance of data offloading to the central cloud and its capability of low-delay edge computing.
In MEC-enabled mobile social networks with caching and D2D communication, according to social relationships, a trust-based resource allocation mechanism is devised \cite{Ref57,Ref58,8647548}.
In addition to dynamic computational capability, time-varying wireless channels, and fluctuating cache status, the acquisition of trust values involving direct and indirect observations from other mobile users should also be considered.
To maximize the utility of the network operator, the DRL-based resource allocation and decision-making algorithm is adopted.
Furthermore, by decoupling states from specific actions, the value function and advantage function are derived from the {\it Q}-function decomposition.
After the separate calculation of the above two functions, their combination is performed at the final fully connected layer to reconstruct the {\it Q}-function to robustly estimate the state values.

Attributed to the characteristics of decentralization, anonymity, and trusted mechanism, many computation-intensive blockchain consensus processes and data processing tasks complicate RL in blockchain-empowered MEC.
In \cite{Ref87}, the pricing cost of local execution, task offloading, task mining, task completion reward, and failure penalties of local execution and offloading is minimized.
Since the cost-related actions form a large-scale space, the adaptive genetic algorithm is introduced to generate a limited number of candidate actions.
Then, DRL-based computational offloading is adopted in the MEC system.
To further improve the existing DRL algorithms in terms of convergence, stability, and robustness, \cite{Ref90} proposed a multiagent DRL-based cooperative computation offloading policy in the NOMA-enabled MEC system with the aid of expert strategies, scatter networks, and hierarchical agents.
Simulation results show that compared to state-of-the-art counterparts, the proposed DRL algorithm can achieve a significant reduction in training time while maintaining stable and robust performance in a constantly changing environment.
In \cite{Ref160}, blockchain-empowered FL was proposed to guarantee learning efficiency and data privacy in large-scale IoT scenarios.
To improve communication efficiency, a DRL-based optimization strategy of trusted user scheduling and spectrum allocation is developed.

Owing to network dynamics and different miners' preferences toward the tradeoff between risks (no profit or excessive payment) and rewards, computation- and communication-constrained resource pricing and allocation are critical issues in the MEC-enabled blockchain.
Meanwhile, the miner (mobile device) always makes the independent decision to maximize its individual payoff without knowing other miners' actions, which is not beneficial to achieve the global solution.
In \cite{Ref88}, a noncooperative game framework was devised for interactions among miners.
Based on the POMDP, the Bayesian RL-based offloading algorithm is developed to reach a perfect Bayesian equilibrium among all miners.
To reduce the RL complexity, Bayesian DRL is adopted by using a Bayesian neural network to reach a perfect Bayesian equilibrium among the short-sighted miners.
Based on the POMDP model for miners in \cite{Ref88}, the long-term payoff of a service provider is considered in \cite{Ref89}, where the decision-making of the service provider is modeled as the MDP.
The interaction between the service provider and miners is modeled as a stochastic Stackelberg game.
The service provider assigns the price and miners select their hash rates.
To maximize the long-term payoff of the service provider and miners, a hierarchical RL is designed by combining the RL-based offloading strategy for miners with the RL-based price assignment for the service provider.
In \cite{9385791}, to maximize the system utility of all miners, A3C-based resource allocation and pricing were proposed.
The system utility is defined as the difference in the price-based probability minus the sum of the transmission price and computational price.
To guarantee the QoE and QoS of blockchain-based MEC in time-varying wireless channels with limited bandwidth resources, adaptive task offloading and resource allocation were investigated in \cite{Ref132}.
With the aid of the double dueling DQN algorithm, the joint optimization problem of spectrum allocation, block size, and number of successive blocks is effectively solved.
Owing to the intermittent connectivity caused by time-varying wireless channels and dynamic processing queues at MEC servers, cooperative computation offloading and power allocation among multiple MEC servers was proposed for blockchain-empowered MEC in \cite{Ref133}.
Cooperative offloading includes direct offloading to a BS and indirect offloading through a relay node to the same BS.
To maximize the computation rate and the transaction throughput, the computation offloading, power control, block size, and block interval are jointly optimized by the A3C-based RL algorithm.
The studies on trust-based applications are shown in Table \ref{tab:tab9}.

\begin{table*}[thpb]
\centering
\caption{Comparisons of RL-Based MEC in the Trust-Based Applications}
\label{tab:tab9}
\begin{tabular}{|m{0.2in}|m{0.65in}|m{0.5in}|c|m{0.3in}|m{0.5in}|m{0.75in}|m{0.5in}|m{0.6in}|m{0.3in}|m{0.6in}|}
\hline
 \multirow{3}{*}{\textbf{Ref.}} & \multicolumn{1}{c|}{\multirow{3}{*}{\textbf{Objective}}} & \multirow{3}{*}{\textbf{Constraints}} & \multirow{3}{*}{\textbf{Sol.}} & \multicolumn{7}{c|}{\textbf{Reinforcement Learning}} \\  \cline{5-11}
   &    & & & \multirow{2}{*}{\textbf{Model}} & \multirow{2}{*}{\textbf{Algorithm}} & \multicolumn{3}{c|}{\textbf{States}} & \multirow{2}{*}{\textbf{Action}} & \multicolumn{1}{c|}{\multirow{2}{*}{\textbf{Reward}}}  \\ \cline{7-9}
    &   &  &   & & & \multicolumn{1}{c|}{\textbf{Mobility}} & \multicolumn{1}{c|}{\textbf{Channel}} & \multicolumn{1}{c|}{\textbf{MEC}} &  &   \\ \hline
  \cite{Ref57,Ref58,8647548}    &   Trust value, resources of computation, communication, caching  &  \multicolumn{1}{c|}{-----}    & S1  & MDP & DRL   &  Trust value, computational resource, content and its version & CSI & Trust value, computational resource, content and its version & \ding{172} \ding{175} \ding{184} & Earnings of communication, computation, caching   \\ \hline
  \cite{Ref87}   &  Processing cost, mining reward, failure penalty, completion reward  &   \multicolumn{1}{c|}{-----}    & S1  & MDP &  Model-free DRL   &  Task queue, CPU frequency, network ID, remaining tokens, brief information & Transmis-sion rate & Price, CPU frequency, task queue, hash power &  \ding{172}   & Negative system cost \\ \hline
      \cite{Ref160}   &  Delay, energy, learning loss   &  MD, CH, MEC   & S2  & \multicolumn{1}{c|}{-----} & FL+DRL   & Loss value, CPU capability, model parameters &  Transmis-sion rate & \multicolumn{1}{c|}{-----}  & \ding{173} \ding{183} &  Sum of negative delay and penalty \\ \hline
   \cite{Ref89}   & Mining reward, computing cost & MD &  S2 & Game, MDP, POMDP   &  RL+RL  &  Reward, price & \multicolumn{1}{c|}{-----} & Budget, action, reward & \ding{180} \ding{174}  & Mining reward, service payoff \\ \hline
  \cite{Ref88}   & Mining reward, computing cost &   \multicolumn{1}{c|}{-----}    & S1 &  Game, POMDP  & Bayesian RL, DRL & Reward & \multicolumn{1}{c|}{-----} & Reward & \ding{175} \ding{182}  & Mining reward and delay penalty \\ \hline
    \cite{Ref90}   & Delay, energy  &  MD, CH, MEC &  S1  &   MDP    &  Multiagent DRL  & Network ID, task size, computational resource & MD's network ID, CSI & Computational resource, decision feedback  & \ding{172} \ding{185}  & Cost of delay and energy consumption \\ \hline
    \cite{Ref132}   &  Delay, throughput   & MD, CH, MEC &  S1 & MDP &  Double dueling DQN  & \multicolumn{1}{c|}{-----} & SNR & Computational resource, network ID  & \ding{183} \ding{176} & Sum of throughput and reciprocal of delay   \\ \hline
      \cite{Ref133}   &  Computation rate, throughput  &  MD, CH, MEC & S1  & MDP & A3C & Trust value  & CSI, stake number & Computational resource & \ding{172} \ding{176} \ding{185}  &  Sum of computation rate and throughput \\ \hline
      \cite{9385791} &  Delay, resource payment & MD, CH, MEC & S1 & MDP & A3C & \multicolumn{1}{c|}{-----}  & SNR, transmission rate, bandwidth & Computational resource & \ding{174} \ding{182} \ding{183}  & Mining reward minus resource payment \\ \hline
\end{tabular}
\end{table*}

\subsection{Learned Lessons}

Diverse network uncertainties exist in multimedia applications, industrial applications, vehicular applications, IoT applications, and trust-based applications.
To solve the optimization problem in a specific application through RL algorithms, blindly capturing the uncertainties is not beneficial to efficiently solving the problem in a targeted manner.
The application type and problem characteristics should be considered, which help to capture the appropriate uncertainties and obtain the results rapidly.
For instance, in multimedia applications, bandwidth consumption is a critical issue in wireless communication.
Thus, the available and consumed bandwidth resources should be mapped into the state space, the bandwidth resource allocation is considered in the action, and the reward is associated with the bandwidth-consuming QoE, such as video quality.
For delay-sensitive industrial applications, the state is strongly related to the computation capabilities of MEC servers and mobile devices and the transmission rate.
Additionally, the action is associated with the computation and transmission resource allocation, and the reward can be defined as a decreasing function of the task completion delay.
Based on the formulated states, actions, and reward in MDP, RL algorithms can be employed for the optimization problem in a specific application.
According to mobile devices' requirements and service deployment, centralized/distributed/hybrid RL can be adopted for communication/computation-efficient learning.
If the state and action spaces are relatively small, the direct solution of S1 can be used.
On the contrary, for large-scale state and action spaces, the decomposition/integration-based solution of S2 is a promising alternative.

\section{Open Challenges}

\subsection{Heterogeneous Resource Management for Diverse Applications}

Diverse applications are always supported by different network functions, each of which needs different kinds and amounts of network resources.
In addition, different applications have different QoS/QoE requirements.
For instance, data-intensive applications, such as movie download, consume many computational resources for high-complexity baseband signal processing (source coding, high-order constellation modulation, multiple-input multiple-output (MIMO) and so on), to achieve high throughput.
Delay-sensitive applications, such as VR and AR, require large amounts of computational and caching resources for application-level processing, such as image rendering and positioning, to satisfy their low-delay requirements.
For deep learning applications, computational resources are consumed for fast and high-accuracy model training.
Moreover, the application/service type and number are large in practical networks, as in \cite{Ref109}.
In this regard, how to allocate heterogeneous resources or slice MEC-based networks to satisfy merged applications should be addressed in an agile and intelligent manner.

\subsection{Sensing Enabled MEC}

Except for the computation, communication, caching, and control, sensing used to collect the data or network environment states will play a key role in the future.
For instance, when computational resources are used to train the model for machine learning, a key problem arises that is where the data come from.
To this end, the interplay between sensing/data collection and computing should be investigated in the MEC.
In addition, integrated sensing and communication has drawn much attention recently and is a promising way to improve sensing and communication efficiency.
The motivation to further integrate sensing, computing, and communication in highly coupled MEC systems remains open.

\subsection{Cooperation Among Distributed Computing Nodes for Distributed Learning}

With the ability to maintain data privacy and improve communication efficiency, distributed learning has become a popular research direction.
Specifically, the MEC-based network can shorten the distance between the computing nodes and the mobile devices, and thus, deploying distributed learning in the MEC-based network is useful for edge intelligence.
For instance, the combination of the FL and the MEC-based vehicular network can learn pedestrian behavior in a distributed manner and help intelligent vehicle decisions.
However, which topology (star topology, hierarchical topology, or mesh topology) is the best for the model interplay among computing nodes should be integrated in the future.

\subsection{Cross-Layer Optimization in MEC}
In \cite{Ref157,8657791}, the offloading decision involving routing was researched with the aid of RL.
In addition, in \cite{Ref59}, an adaptive online adjustment of the initial congestion window was proposed with the goal of minimizing the flow completion time.
The best adjustment policy is achieved by the A3C-based RL algorithm.
To improve the convergence efficiency of the RL algorithm, a bracket of techniques is adopted, including batch normalization, decision reuse of the initial congestion window, and early feedback mechanism.
In this regard, the integration of supervised learning and RL is implemented to achieve a common policy model for dynamic MEC systems.
However, in practical networks, there are multiple network protocols in different layers, such as the transmission control protocol (TCP)  in the transport layer and the internet protocol (IP) in the network layer.
Additionally, in each layer, there are also different protocols, such as in the data link layer including point-to-point protocol (PPP), high-level data link control (HDLC), ALOHA, and carrier sense multiple access (CSMA).
Due to a large number of heterogeneous network protocols in practice, large states and actions are still critical challenges for RL-based MEC.

\subsection{RL in MEC with Multiple Mobility Patterns}
When a mobile device is connected to different network interfaces, such as WiFi and 5G, the energy consumption is different \cite{Ref182}.
Owing to the limited coverage of each MEC server, when a mobile device is going to a long-distance destination, frequent handover through different air interfaces increases the service delay.
Furthermore, when using different transportation, the mobile user will have different mobility patterns, which complicates environmental dynamic predictions.
In this case, with the aid of MEC server cooperation and task migration, managing mobility is a challenge.

In current networks, to make effective decisions and satisfy the requirements of mobile users, network resources should be adaptively allocated.
In contrast, it is interesting that the MEC server is capable of intelligently instructing/controlling mobile users/devices to enjoy better network services.
For instance, APs such as BSs and WiFi are commonly deployed on demand, resulting in different loads of wireless communication, computation, and caching in different regions.
When a mobile user/device goes through different regions with continuous task offloading, if the network can instruct the user to stay a while (or move slower than before) in areas with low loads (e.g., sufficient communication, computation, and caching resources) or if the network can instruct the user to move faster than before in areas with heavy loads (e.g., no available spectrum resource), frequent migration or excessive local computation will be avoided.
Based on mobility shaping for mobile devices/users, the delay/energy consumption of task offloading, content caching, or network control will be efficiently reduced compared to conventional service patterns.
Furthermore, when mobile users' daily routines are considered, how to employ RL to actively recommend more accurate guidance and provide network services is another challenging issue.

\subsection{Backhaul-Constrained MEC}
In wireless networks, the computational tasks are first offloaded to an AP over the wireless channels and then transferred to the MEC servers through backhauls.
In common MEC systems, the transmission cost in the backhauls is always neglected.
However, in the multihop task offloading procedure, the time consumption on the backhaul cannot be ignored anymore \cite{9201170}.
Furthermore, from wireless transmission to limited backhauls (or reverse), due to free mobility, dynamic wireless channels, and distributed MECs, the low-delay requirement of all mobile devices is still challenging in integrated backhaul and wireless MEC-based networks.

\subsection{Green MEC with Sleep Mechanism}

Currently, CO2 emissions are a critical concern worldwide and are closely related to energy consumption.
With the exponential growth of connected mobile devices, extensive computational requirements, and growing energy requirements in higher-frequency communications, energy efficiency is a critical issue in green MEC.
Similar to BSs in cellular networks, a sleep mechanism can be introduced into MEC servers to turn off all or part of MEC functions.
In this way, when an offloading task is requested, one of the critical problems is how to perform server cooperation, computational migration, and MEC server wake-up mechanisms for mobile devices/users.
For example, few people use MEC servers at night, and MEC servers in nonhot areas can enter the hibernation mode.
When MEC servers in hot areas cannot handle all requests, where, when, and how to perform active MEC servers or start dormant MEC servers are critical problems.

\section{Conclusions}

This paper has presented a comprehensive survey of the RL-enabled MEC system in terms of network uncertainty.
First, we have presented a brief overview of MEC-empowered networks and specified the optimization difficulties based on the analysis of the network uncertainties.
Then, we have summarized the RL-based MEC framework, which provides two solutions for network optimization.
The first solution is direct RL, based on MDP models for optimization problems.
The second solution is indirect RL, where based on the decomposition of the original optimization problem into many subproblems, parts of the subproblems are optimized by RL algorithms and others are optimized by conventional optimization algorithms or machine learning algorithms.
Afterward, according to free mobility, time-varying channels, and distributed edge services in heterogeneous MEC-empowered networks, a detailed survey of MEC has been presented using mobility-, spectrum-, computation-, and caching-aware RL.
Finally, potential challenges have been given as future research directions.





\ifCLASSOPTIONcaptionsoff
  \newpage
\fi


\normalem

\bibliographystyle{IEEEtran}
\bibliography{IEEEabrv,ReferenceRLMEC_1_Abrv} 

%



\end{document}